\documentclass[twocolumn, floatfix, superscriptaddress]{revtex4-1}
\usepackage[utf8]{inputenc}
\usepackage{amsfonts}
\usepackage{amsmath}

\usepackage{tabularx}

\usepackage[dvipsnames]{xcolor}
\usepackage{graphicx}
\usepackage{tkz-graph}

\usepackage[colorlinks=true,bookmarks=false,linkcolor=RoyalBlue,urlcolor=RoyalBlue,citecolor=RoyalBlue,breaklinks]{hyperref}

\begin{document}

\title{Emergent entanglement structures and self-similarity in quantum spin chains}

\author{Boris Sokolov}
\affiliation{QTF Centre of Excellence, Turku Centre for Quantum Physics, Department of Physics and Astronomy, University
of Turku, FI-20014 Turun Yliopisto, Finland}

\author{Matteo A.~C.~Rossi}
\affiliation{QTF Centre of Excellence, Turku Centre for Quantum Physics, Department of Physics and Astronomy, University
of Turku, FI-20014 Turun Yliopisto, Finland}

\author{Guillermo Garc\'ia-P\'erez}
\affiliation{QTF Centre of Excellence, Turku Centre for Quantum Physics, Department of Physics and Astronomy, University
of Turku, FI-20014 Turun Yliopisto, Finland}
\affiliation{Complex Systems Research Group, Department of Mathematics and Statistics,
University of Turku, FI-20014 Turun Yliopisto, Finland}

\author{Sabrina Maniscalco}
\affiliation{QTF Centre of Excellence, Turku Centre for Quantum Physics, Department of Physics and Astronomy, University
of Turku, FI-20014 Turun Yliopisto, Finland}
\affiliation{QTF Centre of Excellence, Center for Quantum Engineering, Department of Applied Physics,
Aalto University School of Science, FIN-00076 Aalto, Finland}

\begin{abstract}
    We introduce an experimentally accessible network representation for many-body quantum states based on entanglement between all pairs of its constituents. We illustrate the power of this representation by applying it to a paradigmatic spin chain model, the XX model, and showing that it brings to light new phenomena. The analysis of these entanglement networks reveals that the gradual establishment of quasi-long range order is accompanied by a symmetry regarding single-spin concurrence distributions, as well as by instabilities in the network topology. Moreover, we identify the existence of emergent entanglement structures, spatially localised communities enforced by the global symmetry of the system that can be revealed by model-agnostic community detection algorithms. The network representation further unveils the existence of structural classes and a cyclic self-similarity in the state, which we conjecture to be intimately linked to the community structure. Our results demonstrate that the use of tools and concepts from complex network theory enables the discovery, understanding, and description of new physical phenomena even in models studied for decades.
\end{abstract}

\maketitle

The last few decades, we have witnessed a paradigm shift in one of the most fundamental physical theories ever developed: quantum physics \cite{adesso2018}. This paradigm shift was not just motivated by the realisation that bizarre quantum effects are also powerful resources for technological applications but, more broadly, it was a gradual change in the perception of the range of applicability of quantum theory itself. While initially considered a theory describing the microscopic realm, it is nowadays clear that quantum theory also has an impact and non-negligible consequences in the macroscopic realm, to the point that quantum coherence seems to be present even in hot and noisy simple biological systems \cite{Caoeaaz4888}.

In a sense, the need to bridge the gap between the microscopic and macroscopic physical descriptions of reality, arising from the difference in the laws governing large many-body systems and those of their individual components, has been regarded, since the very birth of quantum theory, as both  a crucial missing ingredient and a most problematic issue \cite{Briggs2013}. The quantum measurement problem, or the quantum-to-classical transition, can be clearly seen as an early example of this issue.

It is only very recently, however, that experiments have been achieving a two-fold feat. On the one hand, they have dramatically increased precision and efficiency of coherent manipulation and measurement of individual quantum systems that are part of large suitably engineered many-body systems. On the other hand, they have been able to perform quantum simulations of such larger systems, e.g., condensed matter systems, under very ``clean'' and controllable conditions \cite{Johnson2014,Zhang2017,Bernien2017,Lloyd1996,Berges2019,Smith2019}. Moreover, the increase in (classical) computational power, and the development of efficient algorithms, have enabled to investigate numerically the properties of larger many-body quantum systems \cite{Montangero2018}. Finally, skilful techniques for tomographic reconstruction of both quantum states and channels have been developed \cite{Banaszek2013,Eisert2020}, together with a variety of error mitigation approaches. This means that we are starting to have at our disposal vast experimental and numerical data sets containing an enormous amount of information on the behaviour of quantum many-body systems. A very relevant question is therefore: how do we analyse and extract as much information as possible from these data?

The impact of this question is evident when noticing that relatively new fields, such as quantum biology and quantum thermodynamics, as well as more established fields, like quantum chemistry and quantum gravity, are nowadays exploring at a deeper level the emergence of complex collective structures, behaviours, and phenomena in systems formed by a large number of individual (interacting) quantum systems. This is also relevant for technological applications, from quantum simulators aimed at, e.g., investigating new drugs or designing new materials, to the scaling up of quantum computers and the quantum internet. Under the light of these considerations, it appears clear that the key conceptual ingredient underpinning the development of modern quantum physics, and we dare saying of modern physics at large, is the concept of ``emergence'', i.e., emergent phenomena and emergent behaviour \cite{Anderson1972}. There exist many definitions of emergence, and we will not dwell here on the related philosophical debate. It suffices to say that by emergent properties we refer to properties that cannot in principle be reduced to or derived from those of the lower-level constituents composing the many-body (complex) system.

For classical complex systems, the development of complex network theory, consequent and motivated by the availability of big data sets, has not only provided a theoretical framework to analyse emergent phenomena but, most importantly, has permitted to introduce models explaining their origin. In the quantum realm, however, a similar step has not yet been undertaken, despite the birth and rise of quantum machine learning \cite{Dunjko2020}. We note, indeed, that although machine learning approaches are commonly associated with enormous predictive power in big data scenarios, they oftentimes lack descriptive power.

We argue that a rigorous way to access and formalise emergent properties and dynamics in quantum complex systems is to merge and, when needed, generalise the approaches and mathematical tools of complex network theory and quantum physics. This amounts to developing the theory of complex quantum networks. Some attempts have been initiated to pursue such a programme \cite{Biamonte2019}. Most of the examples studied fall into two categories: networks of entanglement, wherein connections (links) represent entangled states \cite{Acin2007,Cuquet2009,Perseguers2010}, and networks of quantum systems where the links are physical interactions \cite{Faccin2013,Paparo2012,SanchezBurillo2012,Faccin2014,omar16,Mulken2016,Cabot2018,Nokkala2016,Nokkala2018,Nokkala2018a}. Only very recently, however, the idea of using a network representation to describe the properties of complex many-body quantum states has been put forward \cite{Valdez2017,Bhuvanesh2018}. The latter is the framework we are interested in here, which, importantly, is markedly different from the idea of graph or cluster state used in quantum information theory \cite{Nielsen2006}.

In this paper we establish a crucial step in the description of many-body quantum states as complex networks by proving for the first time that the use of this representation, with the annexed theoretical toolbox, can reveal new emergent phenomena even in extensively studied paradigmatic critical quantum spin chain models.

\begin{table}[t]
\begin{ruledtabular}
\begin{tabular}{p{3cm} l r}
 Degree & $\displaystyle d_i = \sum_{j=1}^N a_{ij} $ &  \begin{tikzpicture}[baseline=2.5ex,auto,scale=.5]
        \clip (-.3,-.3) rectangle (2.72, 2.3);
        \tikzstyle{main}=[shape=circle,red,thick,draw, inner sep = 0.05cm]
        \tikzstyle{others}=[shape=circle,thick,draw, inner sep = 0.05cm]
        \node (top) at ( 0.1,2) [others,draw] {};
        \node (left) at ( 2.42,1) [others,draw] {};
        \node (mid) at ( 1,1) [main,draw] {};
        \node (bottom) at ( 0.1,0) [others,draw] {};
        \draw[thick]   (top) -- (mid)
        (bottom) -- (mid)
        (left) -- (mid);
    \end{tikzpicture} \\
    \hline
    Strength & $\displaystyle s_i = \sum_{j=1}^N \omega_{ij} $ &

    \begin{tikzpicture}[baseline=2.6ex,auto,scale=.5]
        \clip (-.3,-.3) rectangle (2.72, 2.3);
        \tikzstyle{main}=[shape=circle,red,thick,draw, inner sep = 0.05cm]
        \tikzstyle{others}=[shape=circle,thick,draw, inner sep = 0.05cm]
        \node (top) at ( 0.1,2) [others,draw] {};
        \node (left) at ( 2.42,1) [others,draw] {};
        \node (mid) at ( 1,1) [main,draw] {};
        \node (bottom) at ( 0.1,0) [others,draw] {};
        \draw[thin, draw=black!20] (top) -- node  {\scriptsize .2} (mid) ;
        \draw[thick, draw=black!100] (bottom) -- node [swap] {\scriptsize 1} (mid) ;
    \draw[thin, draw=black!60] (left) -- node {\scriptsize .6} (mid) ;
    \end{tikzpicture}\\
    \hline
    Clustering & $\displaystyle c_i = \frac{\sum_{j,k} a_{ij} a_{ik} a_{jk}}{d_i(d_i-1)}$ &
    \begin{tikzpicture}[baseline=2.6ex,auto,scale=.5]
        \clip (-.3,-.3) rectangle (2.72, 2.3);
        \tikzstyle{main}=[shape=circle,red,thick,draw, inner sep = 0.05cm]
        \tikzstyle{others}=[shape=circle,thick,draw, inner sep = 0.05cm]
        \node (top) at ( 0.1,2) [others,draw] {};
        \node (left) at ( 2.42,1) [others,draw] {};
        \node (mid) at ( 1,1) [main,draw] {};
        \node (bottom) at ( 0.1,0) [others,draw] {};
        \draw[thick]   (top) -- (mid)
                (bottom) -- (mid)
                (left) -- (mid)
                (top) -- (bottom);
    \end{tikzpicture} \\
    \hline
    Weighted clustering & $\displaystyle c_i^{\omega} = \frac{\sum_{j,k} ( \omega_{ij} \omega_{ik} \omega_{jk})^{1/3}}{d_i(d_i-1) \max\limits_{lm} \omega_{lm}} $ &
    \begin{tikzpicture}[baseline=2.6ex,auto,scale=.5]
        \clip (-.5,-.3) rectangle (2.72, 2.3);
        \tikzstyle{main}=[shape=circle,red,thick,draw, inner sep = 0.05cm]
        \tikzstyle{others}=[shape=circle,thick,draw, inner sep = 0.05cm]
        \node (top) at ( 0.1,2) [others,draw] {};
        \node (left) at ( 2.42,1) [others,draw] {};
        \node (mid) at ( 1,1) [main,draw] {};
        \node (bottom) at ( 0.1,0) [others,draw] {};
        \draw[thin, draw=black!20] (top) -- node  {\scriptsize .2} (mid) ;
        \draw[thick, draw=black!100] (bottom) -- node [swap] {\scriptsize 1} (mid) ;
        \draw[thin, draw=black!60] (left) -- node {\scriptsize .6} (mid) ;
        \draw[thin, draw=black!40] (top) -- node [swap] {\scriptsize .4} (bottom) ;
    \end{tikzpicture}
    \\ \hline
    Disparity & $\displaystyle Y_i = \frac{1}{s_i^2} \sum_{j=1}^N (\omega_{ij})^2 $ &
    \begin{tikzpicture}[baseline=2.6ex,auto,scale=.5]
        \clip (-.3,-.3) rectangle (3.0, 2.3);
        \tikzstyle{main}=[shape=circle,red,thick,draw, inner sep = 0.05cm]
        \tikzstyle{main2}=[shape=circle,blue,thick,draw, inner sep = 0.05cm]
        \tikzstyle{others}=[shape=circle,thick,draw, inner sep = 0.05cm]
        \node (top) at ( 0,2) [others,draw] {};
        \node (left) at ( 1.7,1) [main2,draw] {};
        \node (mid) at ( 1,1) [main,draw] {};
        \node (bottom) at ( 0,0) [others,draw] {};
        \node (topl) at ( 2.7,2) [others,draw] {};
        \node (bottoml) at ( 2.7,0) [others,draw] {};
        \draw[thin, draw=black!20] (top) -- node  {\scriptsize .2} (mid) ;
        \draw[thick, draw=black!100] (bottom) -- node [swap] {\scriptsize 1} (mid) ;
        \draw[thick, draw=black!60] (left) -- node {\scriptsize .6} (mid) ;
        \draw[thick, draw=black!60] (left) -- node  {\scriptsize .6} (topl) ;
        \draw[thick, draw=black!60] (left) -- node [swap] {\scriptsize .6} (bottoml) ;
    \end{tikzpicture}\\
\end{tabular}
\end{ruledtabular}
\caption{\textbf{Overview of the local network measures used in this paper.} For each measure, we include its definition and a small depiction to illustrate the concept. In the mathematical expressions, $\omega_{ij}$ is the weight of the link between nodes $i$ and $j$ (which we identify with the concurrence for our network representation of quantum states), and $a_{ij}$ are the elements of the adjacency matrix of the network, fulfilling $a_{ij} = \Theta(\omega_{ij})$, where $\Theta(x)$ stands for the Heaviside function. Hence, degree and strength account for the number of connections and total weight of a given node, respectively. The red nodes in their respective illustrations have degree $d = 3$ and $s = 1.8$. The clustering coefficient accounts for the fraction of pairs of neighbours of the node that are connected. In the figure, the red node has $c = 1 / 3$. The weighted version of the clustering used here weights the contribution of each triangle by the geometric mean of the values of the three links involved, normalised by the largest weight in the network, $\max_{lm} \omega_{lm}$; in the example graph, the red node has $c^{\omega} \approx 0.14$. The dispariy $Y_i$ quantifies the heterogeneity of the distribution of the weights of the connections of the node. If all its connections have equal weights $\omega = s_i/d_i$, $Y_i = 1/d_i$ (as for the blue node). Instead, if one of the links dominates, the disparity approaches 1. For the red node, $Y \approx 0.43$.}
\label{tab:networkmeasures}
\end{table}

\section{Complex Quantum Network representation}

In this section, we introduce the quantum network representation of a many-body quantum state, focusing specifically on $N$-qubit systems. Renowned examples of such states are ground states of quantum spin chains and lattices, which are cornerstone models of condensed matter physics. Many relevant physical properties of these states can be inferred from two-body --- or pairwise --- quantities, such as correlators of the form $\langle \sigma_i^l \sigma_j^m\rangle$, where $\sigma_i^l$ and $\sigma_j^m$ are Pauli operators with $i, j \in \lbrace x, y, z \rbrace$ ($l$ and $m$ are spin indices), or quantifiers of bipartite entanglement like concurrence~\cite{DeChiara2018,Amico2008}. Importantly, efficient techniques for performing two-body tomography have been very recently discovered, making all these pairwise quantities experimentally accessible even for large $N$~\cite{cotler2019quantum,pairwisetomo,bonetmonroig2019nearly}. Furthermore, limiting our attention to pairwise quantities naturally leads to a complex network description of the quantum state, and consequently allows us to borrow tools and techniques from classical complex network theory for studying quantum many-body systems. In what follows, we introduce the key network-theoretical concepts required to understand and motivate the rest of the paper.

\begin{figure*}[t!]
    \centering
    \includegraphics[width=\textwidth]{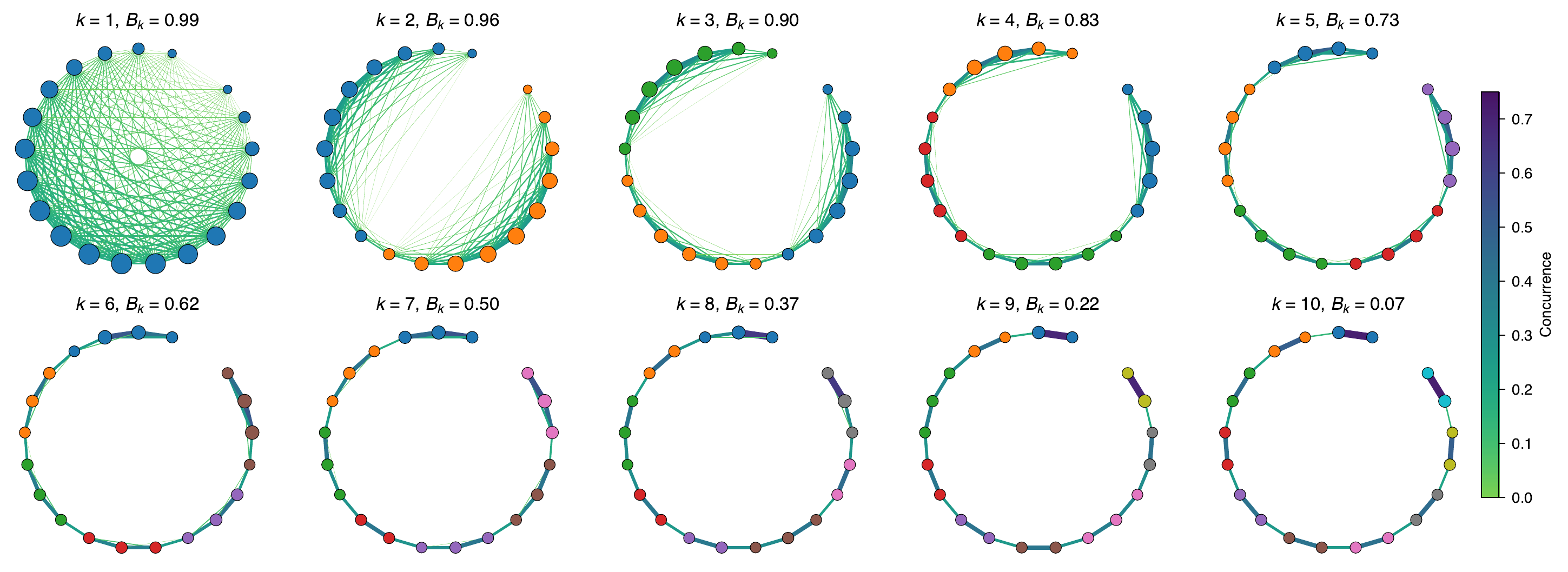}
    \caption{Concurrence networks for $N=20$ spins for different values of the magnetic field $B_k$ for $k = 1, 2, \ldots, 10$. Each node represents a different spin. The width and colour of the links indicate the value of the concurrence between the corresponding pair, while the sizes of the nodes are proportional to their strength. The colours of the nodes identify the community structure detected by the LPA algorithm (cf. Sec.~\ref{sec:Communities}). The number of communities is found to be equal to $k$.}
    \label{fig:concurrence_networks}
\end{figure*}

A complex network is a representation of a complex system in terms of a graph, in which nodes symbolise the individual components of the system and the links represent interactions, correlations, or some other form of relationship between them~\cite{Barrat:2008, Dorogovtsev:2008, Newman:2010}. Notable examples include the internet at the autonomous systems level~\cite{Boguna:2010}, the connectivity among regions of the human brain (or connectome)~\cite{Sporns:2005, Allard:2020}, and social relationships in online social networks~\cite{Kleineberg:2014}. Complex networks are the objects of study of network science, which addresses the description and modelling of their structure~\cite{Garcia-Perez:2018} (or ``connectivity patterns''), as well as how such structure affects the global behaviour of the system (e.g., the spreading of a disease in human contact networks~\cite{Boguna:2003}). Therefore, network theory provides a holistic view of the system under scrutiny, as it studies properties beyond those that can be deduced from or reduced to the ``microscopic'' rules driving each individual part of it. In this paper, we use a network representation for quantum many-body states in which the nodes are spins and the weighted links (i.e., with an associated positive real number) correspond to the pairwise entanglement between them, quantified in terms of concurrence. We show some examples of these networks for ground states of the XX model in Fig.~\ref{fig:concurrence_networks}.

Providing a useful description of a network of even a modest number of nodes (e.g., hundreds) can be a challenging task because of the immense amount of degrees of freedom that graphs possess. Generally, a visual depiction of the network, like those in Fig.~\ref{fig:concurrence_networks}, is useful for small systems only, so one usually must resort to the quantitative study of different relevant properties of the graph. Local measures quantify aspects of the connectivity in the neighbourhood of a specific node. For instance, the \textit{degree} of a node $i$, $d_i$, is the number of links intersecting it, whereas its \textit{strength} $s_i$ is given by the sum of the weights of those links \cite{Newman:2010}. Notice that, while both quantities are similar, the degree only takes into account the existence of a link, regardless of its weight, so it quantifies a property of the unweighted structure of the network, also referred to as the \textit{topology} of the graph in common complex-network parlance. The strength, on the other hand, quantifies a property of the \textit{weighted structure}. The \textit{disparity} of a node $i$, $Y_i$, quantifies the heterogeneity of the distribution of the weights intersecting it \cite{Serrano6483}. The \textit{local clustering coefficient} $c_i$, as well as its weighted generalisation $c_i^\omega$, characterise the density of links among the neighbours of the node \cite{Newman:2010,Onnela2005}. The mathematical definition of each quantity can be found in Table~\ref{tab:networkmeasures}.

The distributions of these quantifiers in the network, or the relations between them, often provide a tangible and understandable, albeit limited, description of even very large systems. In this paper, we will also consider the \textit{community structure} --- a so-called mesoscopic property --- of the entanglement networks. This refers to a property commonly observed in complex networks, namely, the fact that nodes can be grouped into communities such that the density of links among the members of each community is considerably higher than the overall density in the network.

\section{The XX model}

In the literature of quantum spin Hamiltonians, and generally when studying quantum phase transitions, one often works in the thermodynamic limit or considers closed boundary conditions wherein translational invariance is generally guaranteed. This implies that most two-spin correlation functions, including concurrence, which is built upon them, depend only on the distance between the spins \cite{Osterloh2002,Osborne2002,Amico2008}. However, for realistic experimental scenarios, i.e., for quantum simulators such as those realised with trapped ion systems \cite{Lanyon2017} or cold atoms in optical lattices \cite{Gross2017}, the quantum systems are neither close to the thermodynamic limit nor have closed boundary conditions. Therefore, the analysis of the full network of pairwise correlations becomes essential. This is also true, for example, for the study of Majorana fermions and topological defects, where edge and bulk display markedly different properties. This is the reason why we consider the XX model with open boundary conditions in this paper.

We consider a chain of $N$ spins with nearest-neighbours interactions described by the Hamiltonian
\begin{equation}
    H=-J \left[\sum_{i=1}^N \frac{1}{2} (\sigma_x^i \sigma_x^{i+1} + \sigma_y^i \sigma_y^{i+1}) + B \sigma_z ^i\right], \label{eq:XXHamiltonian}
\end{equation}
with $\sigma_{x,y}^{N+1} = 0$, $J$ the coupling constant, which hereafter we set to unity, and $B$ the magnetic field. In the thermodynamic limit, the system undergoes a first order quantum phase transition from a fully polarised state to a critical phase exhibiting quasi-long-range order at $B=1$ \cite{DeChiara2018, Son2009, Amico2008}. The model can be solved exactly by means of Jordan-Wigner transformations \cite{Lieb1961}.

The structure of the ground state and its energy vary with the magnetic field $B$ and, specifically, they depend on a number of level crossings that the system undergoes as $B$ changes. For $B>1$, the ground state energy is $\epsilon_g^0=-NB$ and the ground state, given by $\vert \phi_g
^0 \rangle = \vert \uparrow \rangle^{\otimes N}$, is separable \cite{Son2009}. For $0<B<1$, we can identify $N$ level crossings at values of the magnetic field given by $B_k= \cos[k \pi/(N+1)]$, with $1 \le k \le N$. In each region defined by $B_{k+1} < B < B_k$, the ground state energy is
\begin{eqnarray}
\epsilon_g^k=-(N-2k)B-2\sum_{l=1}^k \cos \left( \frac{\pi l}{N+1}\right)
\end{eqnarray}
and the ground state is
\begin{eqnarray}
|\phi_g^k \rangle = \sum_{l_1<l_2<...<l_k} A_{l_1 l_2 ... l_k} |l_1, l_2, ..., l_k \rangle,
\end{eqnarray}
with $|l_1, l_2, ..., l_k \rangle$ the state with flipped spins at sites $l_1, l_2, ..., l_k $, and amplitudes given by $$A_{l_1 l_2 ... l_k}=\sum_P (-1)^P S_{l_1}^{P(1)}S_{l_2}^{P(2)}...S_{l_k}^{P(k)},$$ where the sum extends over the permutation group of $k$ elements, and where $S_l^k=\sqrt{2/(N+1)} \sin [(\pi k l)/(N+1)]$. At $B=B_k$, the ground state jumps discontinuously from one symmetric subspace to an orthogonal one.

The properties of pairwise concurrence for the XX model were studied in Ref.~\cite{Son2009}, where the  authors used a closed expression for all the reduced two-body density matrices to show that pairwise entanglement presents discontinuous jumps at the transition points $B_k$, and entanglement between two spins in the bulk and at the edge of the chain shows very different behaviour. Specifically, edge entanglement signals the onset of quasi-long range order. We argue that a much more comprehensive view of the properties of the ground state, including new undiscovered features, can be obtained by considering the full pairwise concurrence network.

\section{Bulk, edge, and symmetry near the quantum phase transition}

\begin{figure*}[t!]
    \centering
    \includegraphics[width=\textwidth]{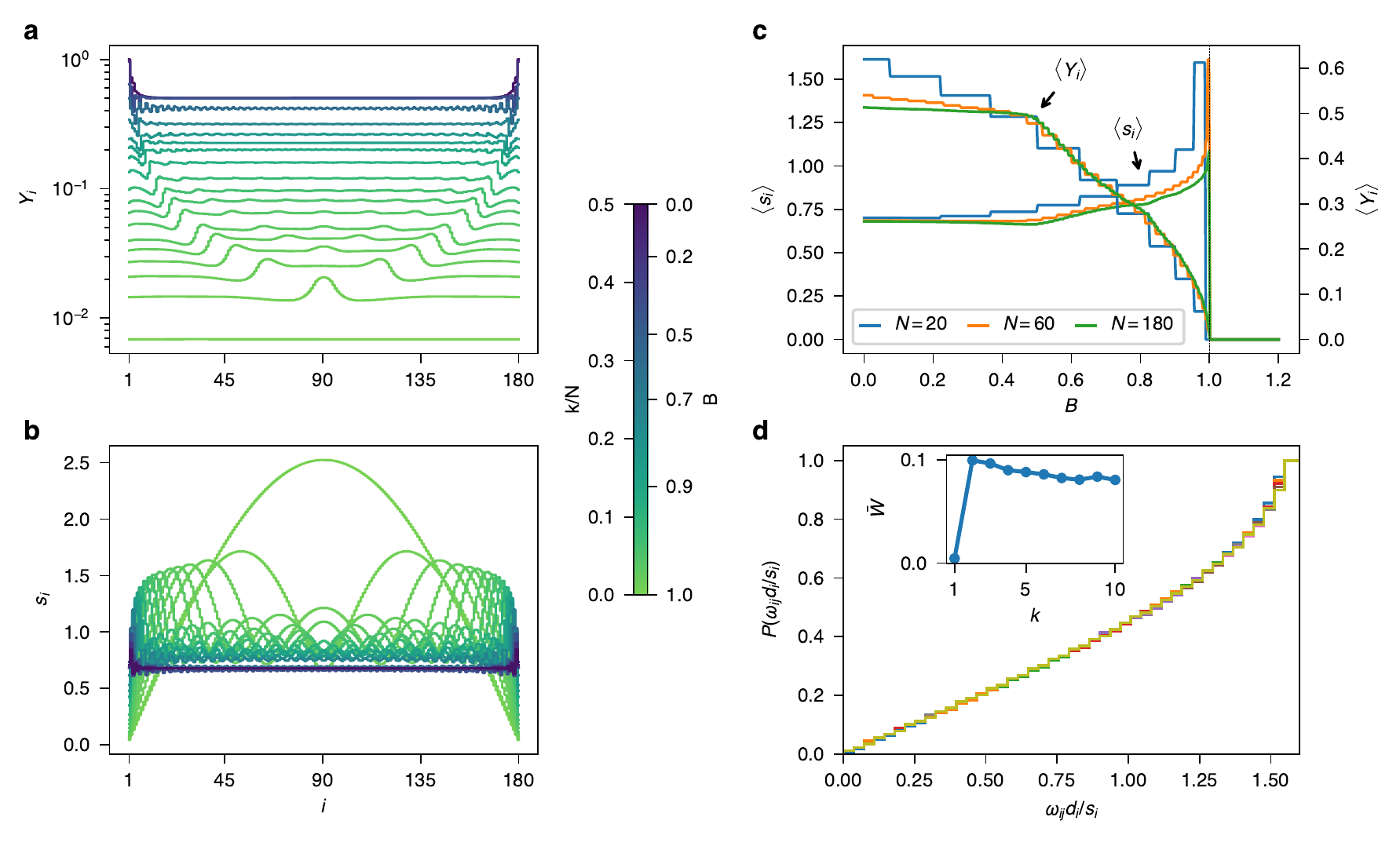}
    \caption{\textbf{Disparity and strength in entanglement networks.} \textbf{a}, Disparity $Y_i$ by node for $N = 180$ and for different values of the magnetic field $0 < B <1$. Note that the disparity has $k-1$ peaks and its average value increases with $k$ (decreases with $B$).
    \textbf{b}, Strength $s_i$ of each node for the same states as in \textbf{a}. The curve corresponding to the $k$-th state presents $k$ maxima. The bar next to the plots indicates the values of the magnetic field $B$ (equivalently, of $k/N$) to which each colour corresponds. For the sake of clarity, the plots do not include the results for all the values of $k$. \textbf{c}, Average strength $\langle s_i \rangle$ and average disparity $\langle Y_i \rangle$ as functions of $B$ for different values of $N$ ($20$, $60$, $180$). \textbf{d}, Each of the overlapping curves depicts the cumulative distribution of the concurrences of a different spin in the chain, rescaled by the average of the distribution, $s_i / d_i$. \textbf{Inset}, Average over all the pairs of spins in the $k$-th state of the Wasserstein distance $W$ between their rescaled local weight distributions. It can be appreciated that this quantity drops to nearly zero for $k = 1$.}
    \label{fig:network_quantities}
\end{figure*}

We start our analysis of the weighted structure of the state by making use of two local measures: strength and disparity, which we depict for each spin in the chain in Fig.~\ref{fig:network_quantities}\textbf{a}-\textbf{b}. It can be appreciated that, close to the critical point $B=1$ ($k=1$), the local (single-spin) distribution of concurrences is very homogeneous and, moreover, the disparity is essentially constant along the chain. At the same time, the strength curve reveals that pairwise entanglement is much stronger for central spins than for those at the ends of the chain. This means, together with the fact that the graph is fully connected (all degrees are equal to $N-1$), that concurrences are actually heterogeneously distributed across the system. Yet, these weights are allocated in such a way that all relative fluctuations at the local level, quantified by the disparity, are equal. This indicates a high level of symmetry in the state right before the quantum phase transition, namely, that the single-spin distributions of concurrence may be very similar for all spins in the chain when appropriately rescaled; this is indeed confirmed in Fig.~\ref{fig:network_quantities}\textbf{d}. From the network point of view, this suggests that the weights can be written as  $\omega_{ij} = \alpha_i \alpha_j$ for some local variables $\lbrace \alpha_i \rbrace$: in such case, the weights of the links intersecting node $i$ rescaled by their mean value, $s_i / d_i$, are $\omega_{ij} / (\sum_{l \neq i} \omega_{il} / d_i)  = (N-1) \alpha_j / \sum_{l \neq i} \alpha_l \approx (N-1) \alpha_j / \sum_{l} \alpha_l$, hence approximately independent of $i$. In fact, in the $k = 1$ concurrence network, $\alpha_i = 2 \sin ( i \pi / (N+1)) / \sqrt{N+1}$.

As the magnetic field is decreased (and $k$ increases), we observe the appearance of $k-1$ peaks in the disparity, signalling a local increase in the heterogeneity of pairwise concurrence for centrally located spins. A close inspection of the plots also shows that the outermost (and highest) peaks in $Y_i$ correspond to the outermost (and lowest) minima in the strength $s_i$, which presents $k$ peaks. In short, we see that there are field-dependent groups of spins near the boundaries exhibiting higher, and more homogeneously distributed, pairwise entanglement, which we may consider as \textit{edges}. The rest of the spins in the chain, with lower and more heterogeneous local concurrence distribution, will be denoted as the \textit{bulk} of the chain. Moreover, these regions are very clearly delimited and their size strongly depends on the magnetic field, for fixed $N$.

Interestingly, as the magnetic field decreases, the difference between the disparity of the bulk of the chain and the one of the edge decreases until, for $B=1/2$  ($k/(N+1) = 1 / 3$) their respective values get inverted, namely the bulk disparity becomes lower than the edge disparity. For very small values of $B$ one observes a pronounced disparity peak for the two outermost spins of the chain, corresponding to their highest value of pairwise concurrence; this phenomenon is a reflection of the fact that, for small $B$, the network has a nearest-neighbour chain topology, as a result of which the edge spins have degree equal to one, and hence their disparity can only be equal to one.

The average behaviour of the network measures is also sensitive to the critical points. This is revealed by the curves of the average disparity $\langle Y_i \rangle$ and the average strength $\langle s_i \rangle$, shown in Fig.~\ref{fig:network_quantities}\textbf{c}, where the discontinuous jumps present for small $N$ gradually become less visible as we approach the thermodynamic limit. The average strength changes discontinuously at the critical point $B=1$, while the average disparity, measuring entanglement heterogeneity, is undefined for $B>1$, since concurrence is zero for all pairs.

\begin{figure*}[t]
    \centering
    \includegraphics[width=\textwidth]{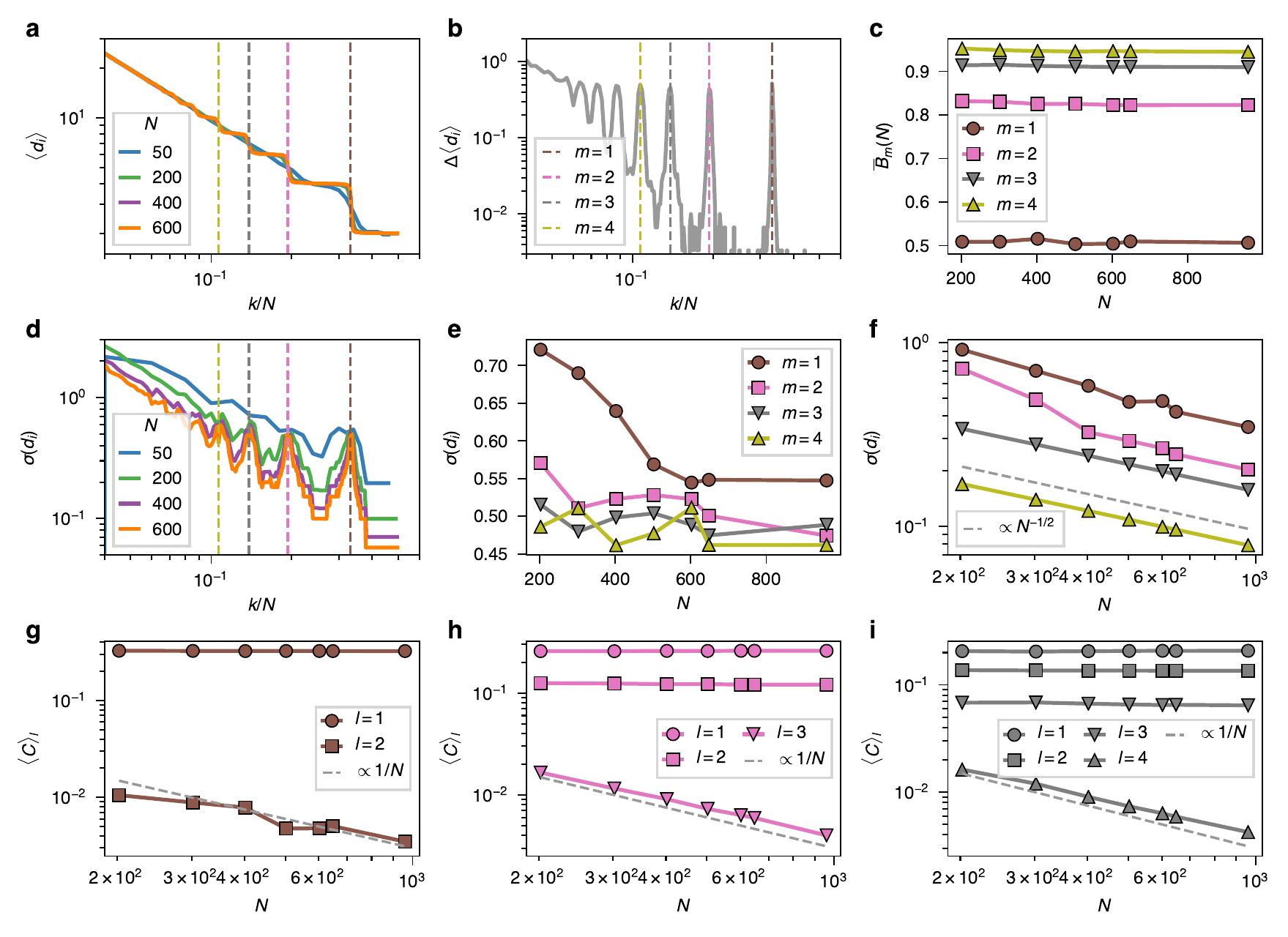}
    \caption{\textbf{Entanglement-topological instabilities.} \textbf{a}, Average degree $\langle d_i \rangle$ as a function of $k / N$ for different system sizes. The change in this quantity becomes increasingly abrupt with increasing system size. \textbf{b}, Change in the average degree $\Delta \langle d_i \rangle$ (calculated via second order central differences) versus $k/N$ for $N = 600$. The vertical lines indicate the values of $k / N$ at which the derivative is maximal. The points are indexed according to $m$, which indicates that the network transitions from an $m$- to an $(m+1)$-nearest neighbours 1D lattice for increasing $B$. \textbf{c}, Position of the four right-most peaks as a function of $N$ in terms of the magnetic field. The transitions remain at fixed values of the field as $N$ increases. \textbf{d}, Standard deviation of the degree $\sigma(d_i)$ of every network as a function of $k / N$ for different system sizes. Notice that the networks become more degree-heterogeneous precisely at the transition points corresponding to the local maxima of the changes of the average degree vs.~$k$ (indicated by vertical dashed lines). \textbf{e}, Scaling of the degree heterogeneity $\sigma(d_i)$ at the four right-most transition points as a function of $N$. \textbf{f}, Scaling of the degree heterogeneity $\sigma(d_i)$ at the midpoints between the five right-most peaks as a function of $N$. The networks become increasingly degree-homogeneous with the system size for values of $B$ different from the transition points (with $\sigma(d_i) \sim N^{-1/2}$, grey dashed line), while they preserve the degree heterogeneity at the peaks. \textbf{g}-\textbf{i}, Average concurrence $\langle C \rangle_l$ of the links of a given length as a function of $N$ at the peaks $m = 1$ \textbf{g}, $m = 2$ \textbf{h}, and $m = 3$ \textbf{i}. At the transitions, the spurious links of length $m+1$ that drive the topology away from a degree-regular graph lose their intensity slightly slower than $1 / N$ (dashed lines). Yet, the links of length up to $m$ at the $m$-th peak remain as $N$ increases.}
    \label{fig:het_at_transition}
\end{figure*}

\section{Entanglement-topological instabilities}
In this section, we turn our attention towards the topological properties of the entanglement networks, that is, the structural properties of the unweighted graphs in which links represent non-separability, disregarding the numerical value of the concurrence when it is non-zero.

We first consider the most elementary network-topological property, the degree $d_i$ (number of connections of a node $i$), and compute its average $\langle d_i \rangle$ over the nodes in the network. Interestingly, as we change $k$, $\langle d_i \rangle$ exhibits abrupt changes at certain level crossings, see Fig.~\ref{fig:het_at_transition}\textbf{a}. For large $N$ and $k$, the average degree only takes even values, $\langle d_i \rangle = 2, 4, 6, \ldots$. Moreover, the curves of $\langle d_i \rangle$ as a function of $k / N$ collapse as $N$ increases, which implies that these sudden transitions occur at specific values of the magnetic field $B$, since the latter is a function of $k / (N+1) \approx k / N$. This is more thoroughly quantified by the results in Figs.~\ref{fig:het_at_transition}\textbf{b}-\textbf{c}. First, we determine the precise values of $k$ at which the transitions take place by identifying the steepest changes in the curves of $\langle d_i \rangle$ versus $k / N$ for every $N$. As an example, in Fig.~\ref{fig:het_at_transition}\textbf{b} we depict the variation of $\langle d_i \rangle$ between consecutive ground states for $N = 600$, indicating four of the identified peaks, labelled with $m$ for increasing $B$, with vertical lines. In Fig.~\ref{fig:het_at_transition}\textbf{c} we show the values of the magnetic field corresponding to each of these peaks, $\bar{B}_m$, as a function of $N$, from which it is clear that they remain constant as $N$ increases.

While the behaviour of the average degree reveals the existence of sudden network-topological transitions, a deeper characterisation of the network structure can be obtained by studying the fluctuations of the degree in every network. In particular, we consider the standard deviation of the degree $\sigma (d_i) = \sqrt{\langle d_i^2 \rangle - \langle d_i \rangle^2}$ as a measure of degree heterogeneity in a network; indeed, $\sigma (d_i) = 0$ if all nodes have the same degree. The behaviour of $\sigma (d_i)$ as a function of $k / N$ reveals that the networks at the transition points $B = \bar{B}_m$ exhibit heterogeneous degree distributions; we name these graphs \textit{entanglement-topological instabilities} (see Fig.~\ref{fig:het_at_transition}\textbf{d}). Furthermore, it can be appreciated that, for $k / N$ different from the values at the peaks --- even in their close neighbourhood --- the degree heterogeneity $\sigma(d_i)$ decreases with increasing system size (see Fig.~\ref{fig:het_at_transition}\textbf{f}), deeming these networks unstable with respect of perturbations in the magnetic field. However, the heterogeneity seems to remain invariant with respect to the increase of system size on the peaks, as indicated by the finite-size scaling analysis in Fig.~\ref{fig:het_at_transition}\textbf{e}.

Our results regarding the off-peak networks are consistent with the findings in Ref.~\cite{Son2009}, where it was shown that, in the thermodynamic limit, concurrence is non-zero only for pairs of spins up to distance $m$, where $m$ depends on the magnetic field. Hence, the resulting networks must be $m$-nearest-neighbours lattices, and all degrees are consequently equal to $2m$. However, the instabilities are characterised by the presence of spurious links of longer length $l \geq m+1$ inhomogeneously distributed across the system, resulting in nodes of different degrees. To conclude our analysis, we show the values of the pairwise concurrence for these spurious links causing the degree heterogeneity in Fig.~\ref{fig:het_at_transition}\textbf{g}-\textbf{i}. We depict the average concurrence of the links of a given length $l$ in the first three instabilities ($m = 1, 2, 3$) as a function of $N$, from which it is clear that the links of length $l \leq m$ do not depend on the system size, while the concurrence of the spurious ones, with $l = m+1$, decreases. In this sense, the phenomenon here discovered is a characteristic of finite size spin chains.

\section{Emergent entanglement structures}
\label{sec:Communities}

\begin{figure}
    \centering
    \includegraphics{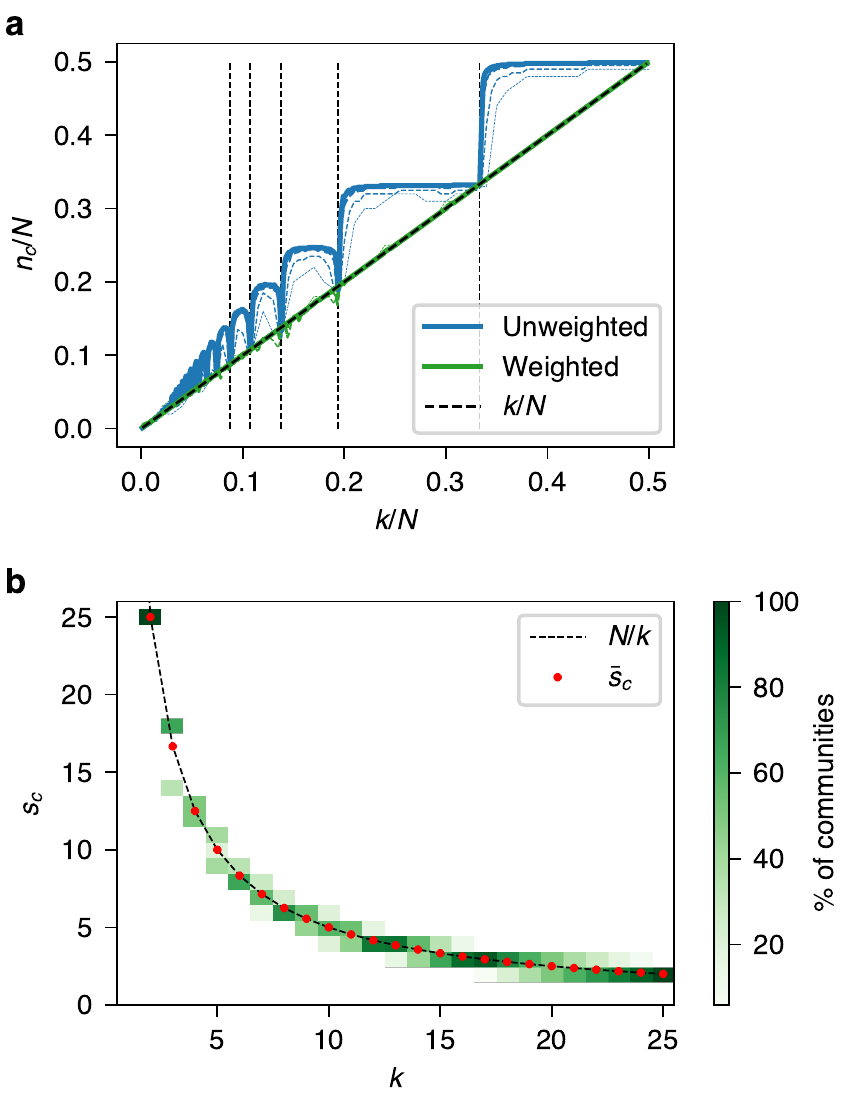}
    \caption{\textbf{Community structure of concurrence networks}. \textbf{a}, Number of communities in the concurrence network over $N$, $n_c / N$, as a function of $k / N$, for $N= 100, 200, 500, 600, 960$, by treating the edges as unweighted (blue) and weighted (green). In the weighted case, the number of detected communities is exactly equal to $k$ (with some fluctuations due to numerical errors), regardless of $N$. In the unweighted case, for increasing $N$ the curve $n_c/N$ vs $k/N$ presents plateaus with dips for the same values of $k/N$ seen in Fig.~\ref{fig:het_at_transition}\textbf{a}-\textbf{b} and \ref{fig:het_at_transition}\textbf{d}. \textbf{b}, Community sizes $s_c$ versus $k$ for $N = 50$. The colour indicates the fraction of communities with a given size. The black dashed line shows $N/k$, while the red points indicate the average size $\bar s_c$ of the detected communities.
    }
    \label{fig:communities}
\end{figure}

The network measures used so far reveal local, i.e. microscopic, structures within the network. It is well-known, however, that networks may also possess mesoscopic structures, which are not uncovered at the level of single nodes, reflecting their behaviour as a whole. Examples are the network community structures describing the heterogeneity in the density of links or in the values of the weights within different subsets of the nodes of the network. We call community a subset of nodes with higher density of connections within the subset than with the rest of the network. The presence of communities is linked to a non-trivial topology of the network: regular and completely random graphs typically do not show any community structure.

Uncovering the community structure of a given graph is an important and computationally demanding task in the analysis of complex networks named \textit{community detection}. Many algorithms have been developed and are being developed with the goal of finding the community structure of large networks quickly and accurately~\cite{Fortunato:2016}. In this paper, we apply a state-of-the-art algorithm, based on label propagation (LPA) \cite{raghavan2007}, which is described in detail in the Methods section.

An example of the communities detected with LPA is shown in Fig.~\ref{fig:concurrence_networks}, where nodes with the same colour belong to the same community. Notice that the algorithm is not provided with any knowledge of the spatial relation among the nodes. In fact, it is completely model-agnostic, in the sense that it is designed to work on arbitrary weighted graphs that may represent any sort of data. Yet, the entanglement structures identified by community detection method have well-defined spatial locations. For small $N$, the community structure is clearly visible from the network representation, as one can see in Fig.~\ref{fig:concurrence_networks}. For small $k$, there are few large communities of nodes with nonzero pairwise entanglement. By increasing $k$, the size of the communities is reduced, up to the limit for $k=N/2$, where all the communities have size $2$, and correspond to pairs of highly entangled spins.

Strikingly, for each value of $k$, the algorithm detects $n_c = k$ communities for any $N$ on the weighted networks, as Fig.~\ref{fig:communities}\textbf{a} shows (there are some small fluctuations, likely due to numerical errors). As a consequence, the average size of the communities is $\bar s_c = N/k$. Moreover, the distribution of community sizes is highly peaked around the mean, as shown in Fig.~\ref{fig:communities}\textbf{b}. It is worth stressing that identifying precisely $k$ communities for large $k$ is especially remarkable considering our description of the network topology in the previous section. Indeed, for low $B$, the graphs are essentially $m$-nearest-neighbours lattices for a wide range of $k$, so the difference in the community structure is far less obvious than the top-left networks in Fig.~\ref{fig:concurrence_networks}. In fact, when fed the unweighted networks (that is, with all the weights transformed as $\omega_{ij} \rightarrow \Theta(\omega_{ij})$), the inferred community structure changes considerably, as can be seen from the blue curves in Fig.~\ref{fig:communities}\textbf{a}. The curves of number of detected communities $n_c$ versus $k$ show plateaus where $n_c$ is roughly constant, interrupted by dips that appear at the same values of $k/N$ for which the average degree changes abruptly (cf.~Fig~\ref{fig:het_at_transition}\textbf{a}-\textbf{b}). Hence, we conclude that, in this regime, the community structure is encoded in the dependence of the concurrence with respect to the position along the chain.

Note that $k$ is the quantum number associated to the total magnetisation $\sum_i \sigma_z^i$, which is a constant of motion. Hence the emergent entanglement structures reflect a global symmetry of the system. Interestingly, following Refs.~\cite{susskind1978,Son2009}, one can describe the properties of the ground state in the dual space by introducing the dual basis $\mu_n= \prod_{m<n} \sigma_x^m$. These operators create a topological excitation (kink) when acting on an initial fully polarised state. As shown in Ref.~\cite{susskind1978}, a single spin flip is equivalent to a kink-antikink pair at two neighbouring dual sites, and any state with $k$ spin flips can be written  as a suitable combination of kink-antikink pairs. In Ref.~\cite{Son2009} the authors notice that, when $B > 1$, the ground state is separable and there are no kinks. Near the critical point $B \lesssim 1$ the ground state consists of a superposition of states with a single spin-flip, or a sea of condensed kink-antikink pairs. This corresponds, in our network representation, to the upper-left network of Fig.~\ref{fig:concurrence_networks}. For smaller and smaller values of $B$, the size of the kink-antikink pairs decreases and their number increases, see the behaviours of the networks of Fig.~\ref{fig:concurrence_networks}, from left to right. At $B = 0$, the ground state has a single kink, with half of the spins pointing down and half pointing up. This phenomenology leads us to conjecture that the emergent entanglement structures that we observe in the spin chain reflect the structure of the kink-antikink pairs in the dual representation described in Refs.~\cite{susskind1978,Son2009}.

\begin{figure*}[t!]
    \centering
    \includegraphics[width=.9\linewidth]{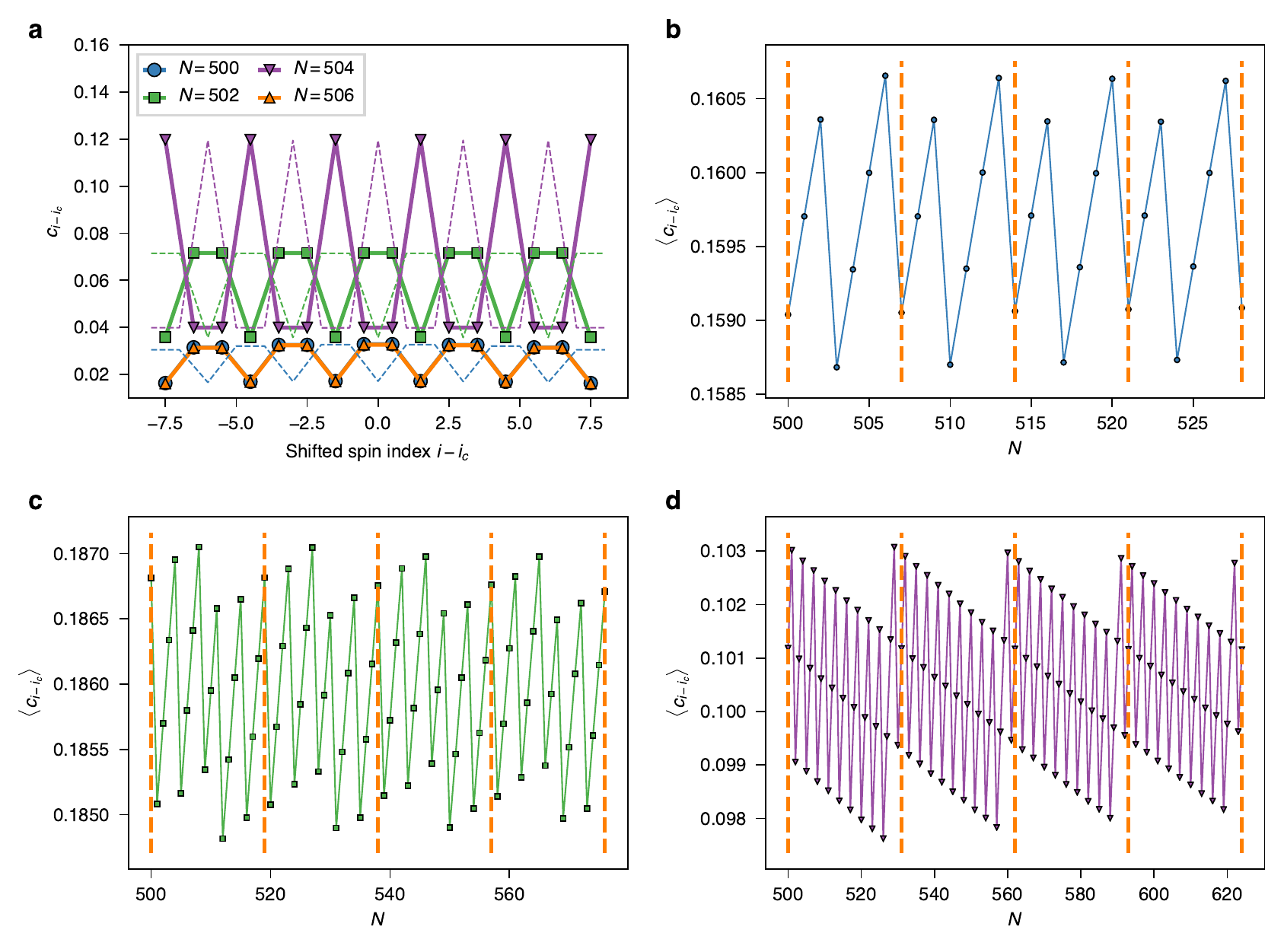}
    \caption{\textbf{Size-periodicity in the network structure.} \textbf{a}, Structural classes revealed by mean local clustering coefficient of nodes at the centre of the spin chain. The indices are shifted so that the central spin (or inter-spin space for even $N$) lies at the centre of the plot. Solid lines with symbol correspond to even $N$, while dashed lines show the profile for odd $N$. Blue, green, and purple dashed lines correspond to $N = 503, 505, 507$, respectively, hence exhibiting the same profile as their even counterparts with equal colour. Moreover, $N=506$ has been included in order to show the nearly perfect overlap with the curve for $N=500$. \textbf{b}-\textbf{d}, Periodicity of the structures quantified through the local clustering averaged over 100 central spins in the chain as a function of $N$. In each plot, the value of the magnetic field was initially fixed as $B = \cos(\pi / s)$ in order to set the expected mean community size to some chosen value $s$, since $s = (N+1)/k \approx \bar{s}_c$. Then, for each value of $N$, we choose $k = \lfloor (N+1) \arccos(B) / \pi \rfloor$ and compute the mean local clustering coefficient for the central spins in the corresponding state. The values of $s$, corresponding magnetic field, and predicted period are: \textbf{b}, $s=3.5$, $B \approx 0.624$, $p = 7$, \textbf{c}, $s=3.8$, $B \approx 0.677$, $p = 19$, \textbf{d}, $s=3.1$, $B \approx 0.529$, $p = 31$. Vertical dashed orange lines are drawn with period $p$ to ease the identification of our heuristic prediction.}
    \label{fig:period}
\end{figure*}

\section{Structural classes and self-similarity}

The network representation also reveals an interesting behaviour of the ground state entanglement structure with respect to changes in the system size. In Fig.~\ref{fig:period}\textbf{a}, we show the weighted clustering coefficient, characterising the density of pairwise entanglement among the neighbours of a node (spin), of 16 central spins in the chain for $N \in [500, 506]$ (17 for odd $N$) at $B = 1/2$. Surprisingly, the addition of a single spin results in a completely different entanglement connectivity pattern in the bulk. What is more, these different patterns exhibit a cyclic self-similarity, in the sense that the addition of 3 spins results in the same clustering profile. As a result, the local connectivity of the concurrence network can be classified into different structural classes. This phenomenon is also observed for other values of the magnetic field with different periods. This is depicted in Fig.~\ref{fig:period}\textbf{b}-\textbf{d}, where we plot the mean local clustering coefficient of the 100 central spins as a function of $N$.

Remarkably, the period of this cyclic behaviour depends on the magnetic field in a way that seems to be closely related to the emergent entanglement structures explored in the previous section. In particular, we conjecture that, if the  mean community size, $\bar{s}_c = N/k$, is a rational number $\bar{s}_c = p / q$, with $p$ and $q$ coprime integers, the period is given by $p$. These periods are indicated with vertical lines in Fig.~\ref{fig:period}\textbf{b}-\textbf{d}. While they perfectly match the period of the mean local clustering in these examples, it should be noted that the periodicity of the structure is not always so clear (e.g., when $p$ is very large). However, understanding whether this is a consequence of considering too small sizes requires further research and will be studied elsewhere.

Our conjecture is greatly motivated by Fig.~\ref{fig:communities}\textbf{b}; more specifically, by the fact that the distribution of community sizes $s_c$ for large $k$ appears to have support mainly over at most two values, $\lfloor \bar{s}_c \rfloor$ and $\lfloor \bar{s}_c \rfloor + 1$ (where $\lfloor x \rfloor$ represents the integer part of $x$) with different weights $1 - f$ and $f$, respectively, such that
\begin{equation}\label{eq:s_c}
    \bar{s}_c = \lfloor \bar{s}_c \rfloor (1-f) + (\lfloor \bar{s}_c \rfloor + 1) f.
\end{equation}
This suggests that the network is organised in groups of $S$ spins such that each group contains an integer number of communities of size $ \lfloor \bar{s}_c \rfloor$ and $ \lfloor \bar{s}_c \rfloor + 1$ in the same proportions as the rest of the network. Hence, one would expect that the addition of precisely $S$ spins would enable the system to organise with a similar network structure. The smallest number of spins $S$ with integer amounts of communities of both sizes in those proportions corresponds to $p$ (see Methods).

\section{Conclusions}
The characterisation and analysis of the properties of many-body quantum states is a daunting task due to the exponential increase in degrees of freedom as the number of particles increases. A novel and, until now, rather unexplored approach is the use of a representation of the quantum state in terms of networks of pairwise quantities, such as entanglement, mutual information, discord, and so on. This representation is particularly useful for systems which do not possess translational invariance, wherein the properties of such correlations do not depend on the distance between the particles only. Moreover, these networks can be obtained efficiently in experimental scenarios, requiring only a logarithmic amount of measurement settings in the system size~\cite{cotler2019quantum,pairwisetomo,bonetmonroig2019nearly}.

In this paper we demonstrate the power of such a representation by applying it to a paradigmatic model of magnetism and showing that it allows us to bring to light new physical characteristics in the ground state of this critical system. Specifically, we unveil (i) a new symmetry in the entanglement structure of the ground state close to the critical point, (ii) the existence of emergent entanglement structures reflecting the global symmetry of the spin Hamiltonian, (iii) the presence of topological instabilities accompanying the onset of quasi-long-range-order, and (iv) self-similarity in the bulk of the pairwise entanglement network.

We have introduced a toolbox of methods and approaches commonly used in complex network theory and demonstrated its usefulness in advancing our knowledge on the complex structure of many-body quantum states. We believe that such a toolbox may prove to be of key importance in the exploration of emergent structures in quantum systems of increasing size and complexity, with implications in quantum biology, quantum simulations, quantum computing, and quantum chemistry.

\begin{acknowledgments}
The authors acknowledge financial support from the Academy of Finland via the Centre of Excellence program (Project no.~312058). G.G.-P.~acknowledges support from the emmy.network foundation under the aegis of the Fondation de Luxembourg. B.S. acknowledges financial support from the Jenny and Antti Wihuri Foundation.
The computer resources of the Finnish IT Center for Science (CSC) and the FGCI project (Finland) are acknowledged.
\end{acknowledgments}

\section*{Author contributions}
All authors contributed equally to the design and implementation of the research, to the analysis of the results, and to the writing of the manuscript.

\section*{Additional information}
{\bf Competing financial interests:} The authors declare no competing financial interests.

\section{Methods}

\subsection{Label propagation algorithm for community detection}
\label{subs:mesoscopic_network_measures}

In this paper, we use a community detection algorithm based on label propagation (LPA) \cite{raghavan2007}. The choice is mainly based on the availability of an open-source implementation, and on the possibility to take the weights into account in the community detection. Specifically, we utilise a semi-synchronous LPA introduced in \cite{cordasco2010} and included in the NetworkX Python package \cite{networkx}. The code has been slightly modified to consider weighted links. We expect other algorithms to give similar results.

The principle of the algorithm is the following: each node is initially assigned a different community label. Then, by following a certain update schedule, the label of a node is changed into the label with the highest frequency among the node's neighbours (optionally weighted using the edge weight). The update schedule in the semi-synchronous algorithm is based on a graph colouring such that no neighbouring nodes have the same colour. All the labels of the vertices belonging to the same colour are updated at once, for all the colours in the network. Ties can occur if there is more than one label with the same highest frequency between the neighbours (this can happen, in particular, when considering unweighted networks). In the case of a tie, the latter is resolved deterministically using a so-called Prec-Max method \cite{cordasco2010}: if the current label of the node is among the labels with highest frequency, then do not change it; otherwise, choose the label with the highest preassigned priority (for example, if each label is an integer number, give priority to the largest number).

\subsection{Smallest spin group with average community structure}

From Eq.~\eqref{eq:s_c}, we see that the fraction of larger communities $f$ is given by the decimal part of $\bar{s}_c$,
\begin{equation}
    f = \bar{s}_c - \lfloor \bar{s}_c \rfloor = (p \mod q) / q.
\end{equation}
The value of $S$ can be computed by considering that the number of spins in communities of size $ \lfloor \bar{s}_c \rfloor + 1$ in each such group, $S \rho_{ \lfloor \bar{s}_c \rfloor + 1}$ (where $\rho_{ \lfloor \bar{s}_c \rfloor + 1} = (\lfloor \bar{s}_c \rfloor + 1) f/[(\lfloor \bar{s}_c \rfloor + 1) f + \lfloor \bar{s}_c \rfloor (1-f)] = (\lfloor \bar{s}_c \rfloor + 1) f/\bar{s}_c$ is the density of spins in communities of size $ \lfloor \bar{s}_c \rfloor + 1$), is a multiple of $ \lfloor \bar{s}_c \rfloor + 1$. In other words, $S \rho_{ \lfloor \bar{s}_c \rfloor + 1} / (\lfloor \bar{s}_c \rfloor + 1) = S f/\bar{s}_c = S (p \mod q)/p$ must be integer. A similar analysis for the communities of size $\lfloor \bar{s}_c \rfloor$ leads to $S (q - p \mod q)/p \in \mathbb{N}$. If $S$ fulfils the first condition, fulfilling the second requires $S q / p \in \mathbb{N}$ and, therefore, $S$ must be a multiple of $p$, given that $p$ and $q$ are coprime.

\bibliography{bibliography.bib}

\begin{thebibliography}{58}%
\makeatletter
\providecommand \@ifxundefined [1]{%
 \@ifx{#1\undefined}
}%
\providecommand \@ifnum [1]{%
 \ifnum #1\expandafter \@firstoftwo
 \else \expandafter \@secondoftwo
 \fi
}%
\providecommand \@ifx [1]{%
 \ifx #1\expandafter \@firstoftwo
 \else \expandafter \@secondoftwo
 \fi
}%
\providecommand \natexlab [1]{#1}%
\providecommand \enquote  [1]{``#1''}%
\providecommand \bibnamefont  [1]{#1}%
\providecommand \bibfnamefont [1]{#1}%
\providecommand \citenamefont [1]{#1}%
\providecommand \href@noop [0]{\@secondoftwo}%
\providecommand \href [0]{\begingroup \@sanitize@url \@href}%
\providecommand \@href[1]{\@@startlink{#1}\@@href}%
\providecommand \@@href[1]{\endgroup#1\@@endlink}%
\providecommand \@sanitize@url [0]{\catcode `\\12\catcode `\$12\catcode
  `\&12\catcode `\#12\catcode `\^12\catcode `\_12\catcode `\%12\relax}%
\providecommand \@@startlink[1]{}%
\providecommand \@@endlink[0]{}%
\providecommand \url  [0]{\begingroup\@sanitize@url \@url }%
\providecommand \@url [1]{\endgroup\@href {#1}{\urlprefix }}%
\providecommand \urlprefix  [0]{URL }%
\providecommand \Eprint [0]{\href }%
\providecommand \doibase [0]{https://doi.org/}%
\providecommand \selectlanguage [0]{\@gobble}%
\providecommand \bibinfo  [0]{\@secondoftwo}%
\providecommand \bibfield  [0]{\@secondoftwo}%
\providecommand \translation [1]{[#1]}%
\providecommand \BibitemOpen [0]{}%
\providecommand \bibitemStop [0]{}%
\providecommand \bibitemNoStop [0]{.\EOS\space}%
\providecommand \EOS [0]{\spacefactor3000\relax}%
\providecommand \BibitemShut  [1]{\csname bibitem#1\endcsname}%
\let\auto@bib@innerbib\@empty
\bibitem [{\citenamefont {Adesso}\ \emph {et~al.}(2018)\citenamefont {Adesso},
  \citenamefont {Franco},\ and\ \citenamefont {Parigi}}]{adesso2018}%
  \BibitemOpen
  \bibfield  {author} {\bibinfo {author} {\bibfnamefont {G.}~\bibnamefont
  {Adesso}}, \bibinfo {author} {\bibfnamefont {R.~L.}\ \bibnamefont {Franco}},\
  and\ \bibinfo {author} {\bibfnamefont {V.}~\bibnamefont {Parigi}},\
  }\bibfield  {title} {\bibinfo {title} {Foundations of quantum mechanics and
  their impact on contemporary society},\ }\href
  {https://doi.org/10.1098/rsta.2018.0112} {\bibfield  {journal} {\bibinfo
  {journal} {Phil. Trans. R. Soc. A}\ }\textbf {\bibinfo {volume} {376}},\
  \bibinfo {pages} {20180112} (\bibinfo {year} {2018})}\BibitemShut {NoStop}%
\bibitem [{\citenamefont {Cao}\ \emph {et~al.}(2020)\citenamefont {Cao},
  \citenamefont {Cogdell}, \citenamefont {Coker}, \citenamefont {Duan},
  \citenamefont {Hauer}, \citenamefont {Kleinekath{\"o}fer}, \citenamefont
  {Jansen}, \citenamefont {Man{\v c}al}, \citenamefont {Miller}, \citenamefont
  {Ogilvie}, \citenamefont {Prokhorenko}, \citenamefont {Renger}, \citenamefont
  {Tan}, \citenamefont {Tempelaar}, \citenamefont {Thorwart}, \citenamefont
  {Thyrhaug}, \citenamefont {Westenhoff},\ and\ \citenamefont
  {Zigmantas}}]{Caoeaaz4888}%
  \BibitemOpen
  \bibfield  {author} {\bibinfo {author} {\bibfnamefont {J.}~\bibnamefont
  {Cao}}, \bibinfo {author} {\bibfnamefont {R.~J.}\ \bibnamefont {Cogdell}},
  \bibinfo {author} {\bibfnamefont {D.~F.}\ \bibnamefont {Coker}}, \bibinfo
  {author} {\bibfnamefont {H.-G.}\ \bibnamefont {Duan}}, \bibinfo {author}
  {\bibfnamefont {J.}~\bibnamefont {Hauer}}, \bibinfo {author} {\bibfnamefont
  {U.}~\bibnamefont {Kleinekath{\"o}fer}}, \bibinfo {author} {\bibfnamefont
  {T.~L.~C.}\ \bibnamefont {Jansen}}, \bibinfo {author} {\bibfnamefont
  {T.}~\bibnamefont {Man{\v c}al}}, \bibinfo {author} {\bibfnamefont
  {R.~J.~D.}\ \bibnamefont {Miller}}, \bibinfo {author} {\bibfnamefont {J.~P.}\
  \bibnamefont {Ogilvie}}, \bibinfo {author} {\bibfnamefont {V.~I.}\
  \bibnamefont {Prokhorenko}}, \bibinfo {author} {\bibfnamefont
  {T.}~\bibnamefont {Renger}}, \bibinfo {author} {\bibfnamefont {H.-S.}\
  \bibnamefont {Tan}}, \bibinfo {author} {\bibfnamefont {R.}~\bibnamefont
  {Tempelaar}}, \bibinfo {author} {\bibfnamefont {M.}~\bibnamefont {Thorwart}},
  \bibinfo {author} {\bibfnamefont {E.}~\bibnamefont {Thyrhaug}}, \bibinfo
  {author} {\bibfnamefont {S.}~\bibnamefont {Westenhoff}},\ and\ \bibinfo
  {author} {\bibfnamefont {D.}~\bibnamefont {Zigmantas}},\ }\bibfield  {title}
  {\bibinfo {title} {Quantum biology revisited},\ }\href
  {https://doi.org/10.1126/sciadv.aaz4888} {\bibfield  {journal} {\bibinfo
  {journal} {Science Advances}\ }\textbf {\bibinfo {volume} {6}},\ \bibinfo
  {pages} {eaaz4888} (\bibinfo {year} {2020})}\BibitemShut {NoStop}%
\bibitem [{\citenamefont {Briggs}\ \emph {et~al.}(2013)\citenamefont {Briggs},
  \citenamefont {Butterfield},\ and\ \citenamefont {Zeilinger}}]{Briggs2013}%
  \BibitemOpen
  \bibfield  {author} {\bibinfo {author} {\bibfnamefont {G.~A.~D.}\
  \bibnamefont {Briggs}}, \bibinfo {author} {\bibfnamefont {J.~N.}\
  \bibnamefont {Butterfield}},\ and\ \bibinfo {author} {\bibfnamefont
  {A.}~\bibnamefont {Zeilinger}},\ }\bibfield  {title} {\bibinfo {title} {{The
  Oxford Questions on the foundations of quantum physics}},\ }\href
  {https://doi.org/10.1098/rspa.2013.0299} {\bibfield  {journal} {\bibinfo
  {journal} {Proc. R. Soc. A}\ }\textbf {\bibinfo {volume} {469}},\ \bibinfo
  {pages} {20130299} (\bibinfo {year} {2013})}\BibitemShut {NoStop}%
\bibitem [{\citenamefont {Johnson}\ \emph {et~al.}(2014)\citenamefont
  {Johnson}, \citenamefont {Clark},\ and\ \citenamefont
  {Jaksch}}]{Johnson2014}%
  \BibitemOpen
  \bibfield  {author} {\bibinfo {author} {\bibfnamefont {T.~H.}\ \bibnamefont
  {Johnson}}, \bibinfo {author} {\bibfnamefont {S.~R.}\ \bibnamefont {Clark}},\
  and\ \bibinfo {author} {\bibfnamefont {D.}~\bibnamefont {Jaksch}},\
  }\bibfield  {title} {\bibinfo {title} {What is a quantum simulator?},\ }\href
  {https://doi.org/10.1140/epjqt10} {\bibfield  {journal} {\bibinfo  {journal}
  {{EPJ} Quantum Technology}\ }\textbf {\bibinfo {volume} {1}},\ \bibinfo
  {pages} {10} (\bibinfo {year} {2014})}\BibitemShut {NoStop}%
\bibitem [{\citenamefont {Zhang}\ \emph {et~al.}(2017)\citenamefont {Zhang},
  \citenamefont {Pagano}, \citenamefont {Hess}, \citenamefont {Kyprianidis},
  \citenamefont {Becker}, \citenamefont {Kaplan}, \citenamefont {Gorshkov},
  \citenamefont {Gong},\ and\ \citenamefont {Monroe}}]{Zhang2017}%
  \BibitemOpen
  \bibfield  {author} {\bibinfo {author} {\bibfnamefont {J.}~\bibnamefont
  {Zhang}}, \bibinfo {author} {\bibfnamefont {G.}~\bibnamefont {Pagano}},
  \bibinfo {author} {\bibfnamefont {P.~W.}\ \bibnamefont {Hess}}, \bibinfo
  {author} {\bibfnamefont {A.}~\bibnamefont {Kyprianidis}}, \bibinfo {author}
  {\bibfnamefont {P.}~\bibnamefont {Becker}}, \bibinfo {author} {\bibfnamefont
  {H.}~\bibnamefont {Kaplan}}, \bibinfo {author} {\bibfnamefont {A.~V.}\
  \bibnamefont {Gorshkov}}, \bibinfo {author} {\bibfnamefont {Z.-X.}\
  \bibnamefont {Gong}},\ and\ \bibinfo {author} {\bibfnamefont
  {C.}~\bibnamefont {Monroe}},\ }\bibfield  {title} {\bibinfo {title}
  {Observation of a many-body dynamical phase transition with a 53-qubit
  quantum simulator},\ }\href {https://doi.org/10.1038/nature24654} {\bibfield
  {journal} {\bibinfo  {journal} {Nature}\ }\textbf {\bibinfo {volume} {551}},\
  \bibinfo {pages} {601} (\bibinfo {year} {2017})}\BibitemShut {NoStop}%
\bibitem [{\citenamefont {Bernien}\ \emph {et~al.}(2017)\citenamefont
  {Bernien}, \citenamefont {Schwartz}, \citenamefont {Keesling}, \citenamefont
  {Levine}, \citenamefont {Omran}, \citenamefont {Pichler}, \citenamefont
  {Choi}, \citenamefont {Zibrov}, \citenamefont {Endres}, \citenamefont
  {Greiner}, \citenamefont {Vuleti{\'{c}}},\ and\ \citenamefont
  {Lukin}}]{Bernien2017}%
  \BibitemOpen
  \bibfield  {author} {\bibinfo {author} {\bibfnamefont {H.}~\bibnamefont
  {Bernien}}, \bibinfo {author} {\bibfnamefont {S.}~\bibnamefont {Schwartz}},
  \bibinfo {author} {\bibfnamefont {A.}~\bibnamefont {Keesling}}, \bibinfo
  {author} {\bibfnamefont {H.}~\bibnamefont {Levine}}, \bibinfo {author}
  {\bibfnamefont {A.}~\bibnamefont {Omran}}, \bibinfo {author} {\bibfnamefont
  {H.}~\bibnamefont {Pichler}}, \bibinfo {author} {\bibfnamefont
  {S.}~\bibnamefont {Choi}}, \bibinfo {author} {\bibfnamefont {A.~S.}\
  \bibnamefont {Zibrov}}, \bibinfo {author} {\bibfnamefont {M.}~\bibnamefont
  {Endres}}, \bibinfo {author} {\bibfnamefont {M.}~\bibnamefont {Greiner}},
  \bibinfo {author} {\bibfnamefont {V.}~\bibnamefont {Vuleti{\'{c}}}},\ and\
  \bibinfo {author} {\bibfnamefont {M.~D.}\ \bibnamefont {Lukin}},\ }\bibfield
  {title} {\bibinfo {title} {Probing many-body dynamics on a 51-atom quantum
  simulator},\ }\href {https://doi.org/10.1038/nature24622} {\bibfield
  {journal} {\bibinfo  {journal} {Nature}\ }\textbf {\bibinfo {volume} {551}},\
  \bibinfo {pages} {579} (\bibinfo {year} {2017})}\BibitemShut {NoStop}%
\bibitem [{\citenamefont {Lloyd}(1996)}]{Lloyd1996}%
  \BibitemOpen
  \bibfield  {author} {\bibinfo {author} {\bibfnamefont {S.}~\bibnamefont
  {Lloyd}},\ }\bibfield  {title} {\bibinfo {title} {Universal quantum
  simulators},\ }\href {https://doi.org/10.1126/science.273.5278.1073}
  {\bibfield  {journal} {\bibinfo  {journal} {Science}\ }\textbf {\bibinfo
  {volume} {273}},\ \bibinfo {pages} {1073} (\bibinfo {year}
  {1996})}\BibitemShut {NoStop}%
\bibitem [{\citenamefont {Berges}(2019)}]{Berges2019}%
  \BibitemOpen
  \bibfield  {author} {\bibinfo {author} {\bibfnamefont {J.}~\bibnamefont
  {Berges}},\ }\bibfield  {title} {\bibinfo {title} {Scaling up quantum
  simulations},\ }\href {https://doi.org/10.1038/d41586-019-01483-1} {\bibfield
   {journal} {\bibinfo  {journal} {Nature}\ }\textbf {\bibinfo {volume}
  {569}},\ \bibinfo {pages} {339} (\bibinfo {year} {2019})}\BibitemShut
  {NoStop}%
\bibitem [{\citenamefont {Smith}\ \emph {et~al.}(2019)\citenamefont {Smith},
  \citenamefont {Kim}, \citenamefont {Pollmann},\ and\ \citenamefont
  {Knolle}}]{Smith2019}%
  \BibitemOpen
  \bibfield  {author} {\bibinfo {author} {\bibfnamefont {A.}~\bibnamefont
  {Smith}}, \bibinfo {author} {\bibfnamefont {M.~S.}\ \bibnamefont {Kim}},
  \bibinfo {author} {\bibfnamefont {F.}~\bibnamefont {Pollmann}},\ and\
  \bibinfo {author} {\bibfnamefont {J.}~\bibnamefont {Knolle}},\ }\bibfield
  {title} {\bibinfo {title} {Simulating quantum many-body dynamics on a current
  digital quantum computer},\ }\href
  {https://doi.org/10.1038/s41534-019-0217-0} {\bibfield  {journal} {\bibinfo
  {journal} {npj Quantum Inf.}\ }\textbf {\bibinfo {volume} {5}},\ \bibinfo
  {pages} {106} (\bibinfo {year} {2019})}\BibitemShut {NoStop}%
\bibitem [{\citenamefont {Montangero}(2018)}]{Montangero2018}%
  \BibitemOpen
  \bibfield  {author} {\bibinfo {author} {\bibfnamefont {S.}~\bibnamefont
  {Montangero}},\ }\href {https://doi.org/10.1007/978-3-030-01409-4} {\emph
  {\bibinfo {title} {Introduction to Tensor Network Methods}}}\ (\bibinfo
  {publisher} {Springer International Publishing},\ \bibinfo {year}
  {2018})\BibitemShut {NoStop}%
\bibitem [{\citenamefont {Banaszek}\ \emph {et~al.}(2013)\citenamefont
  {Banaszek}, \citenamefont {Cramer},\ and\ \citenamefont
  {Gross}}]{Banaszek2013}%
  \BibitemOpen
  \bibfield  {author} {\bibinfo {author} {\bibfnamefont {K.}~\bibnamefont
  {Banaszek}}, \bibinfo {author} {\bibfnamefont {M.}~\bibnamefont {Cramer}},\
  and\ \bibinfo {author} {\bibfnamefont {D.}~\bibnamefont {Gross}},\ }\bibfield
   {title} {\bibinfo {title} {Focus on quantum tomography},\ }\href
  {https://doi.org/10.1088/1367-2630/15/12/125020} {\bibfield  {journal}
  {\bibinfo  {journal} {New J. Phys.}\ }\textbf {\bibinfo {volume} {15}},\
  \bibinfo {pages} {125020} (\bibinfo {year} {2013})}\BibitemShut {NoStop}%
\bibitem [{\citenamefont {Eisert}\ \emph {et~al.}(2020)\citenamefont {Eisert},
  \citenamefont {Hangleiter}, \citenamefont {Walk}, \citenamefont {Roth},
  \citenamefont {Markham}, \citenamefont {Parekh}, \citenamefont {Chabaud},\
  and\ \citenamefont {Kashefi}}]{Eisert2020}%
  \BibitemOpen
  \bibfield  {author} {\bibinfo {author} {\bibfnamefont {J.}~\bibnamefont
  {Eisert}}, \bibinfo {author} {\bibfnamefont {D.}~\bibnamefont {Hangleiter}},
  \bibinfo {author} {\bibfnamefont {N.}~\bibnamefont {Walk}}, \bibinfo {author}
  {\bibfnamefont {I.}~\bibnamefont {Roth}}, \bibinfo {author} {\bibfnamefont
  {D.}~\bibnamefont {Markham}}, \bibinfo {author} {\bibfnamefont
  {R.}~\bibnamefont {Parekh}}, \bibinfo {author} {\bibfnamefont
  {U.}~\bibnamefont {Chabaud}},\ and\ \bibinfo {author} {\bibfnamefont
  {E.}~\bibnamefont {Kashefi}},\ }\bibfield  {title} {\bibinfo {title} {Quantum
  certification and benchmarking},\ }\href
  {https://doi.org/10.1038/s42254-020-0186-4} {\bibfield  {journal} {\bibinfo
  {journal} {Nat. Rev. Phys.}\ ,\ \bibinfo {pages} {382–390}} (\bibinfo
  {year} {2020})}\BibitemShut {NoStop}%
\bibitem [{\citenamefont {Anderson}(1972)}]{Anderson1972}%
  \BibitemOpen
  \bibfield  {author} {\bibinfo {author} {\bibfnamefont {P.~W.}\ \bibnamefont
  {Anderson}},\ }\bibfield  {title} {\bibinfo {title} {More is different},\
  }\href {https://doi.org/10.1126/science.177.4047.393} {\bibfield  {journal}
  {\bibinfo  {journal} {Science}\ }\textbf {\bibinfo {volume} {177}},\ \bibinfo
  {pages} {393} (\bibinfo {year} {1972})}\BibitemShut {NoStop}%
\bibitem [{\citenamefont {Dunjko}\ and\ \citenamefont
  {Wittek}(2020)}]{Dunjko2020}%
  \BibitemOpen
  \bibfield  {author} {\bibinfo {author} {\bibfnamefont {V.}~\bibnamefont
  {Dunjko}}\ and\ \bibinfo {author} {\bibfnamefont {P.}~\bibnamefont
  {Wittek}},\ }\bibfield  {title} {\bibinfo {title} {A non-review of quantum
  machine learning: trends and explorations},\ }\href
  {https://doi.org/10.22331/qv-2020-03-17-32} {\bibfield  {journal} {\bibinfo
  {journal} {Quantum Views}\ }\textbf {\bibinfo {volume} {4}},\ \bibinfo
  {pages} {32} (\bibinfo {year} {2020})}\BibitemShut {NoStop}%
\bibitem [{\citenamefont {Biamonte}\ \emph {et~al.}(2019)\citenamefont
  {Biamonte}, \citenamefont {Faccin},\ and\ \citenamefont
  {Domenico}}]{Biamonte2019}%
  \BibitemOpen
  \bibfield  {author} {\bibinfo {author} {\bibfnamefont {J.}~\bibnamefont
  {Biamonte}}, \bibinfo {author} {\bibfnamefont {M.}~\bibnamefont {Faccin}},\
  and\ \bibinfo {author} {\bibfnamefont {M.~D.}\ \bibnamefont {Domenico}},\
  }\bibfield  {title} {\bibinfo {title} {Complex networks from classical to
  quantum},\ }\href {https://doi.org/10.1038/s42005-019-0152-6} {\bibfield
  {journal} {\bibinfo  {journal} {Commun. Phys.}\ }\textbf {\bibinfo {volume}
  {2}},\ \bibinfo {pages} {53} (\bibinfo {year} {2019})}\BibitemShut {NoStop}%
\bibitem [{\citenamefont {Ac{\'{\i}}n}\ \emph {et~al.}(2007)\citenamefont
  {Ac{\'{\i}}n}, \citenamefont {Cirac},\ and\ \citenamefont
  {Lewenstein}}]{Acin2007}%
  \BibitemOpen
  \bibfield  {author} {\bibinfo {author} {\bibfnamefont {A.}~\bibnamefont
  {Ac{\'{\i}}n}}, \bibinfo {author} {\bibfnamefont {J.~I.}\ \bibnamefont
  {Cirac}},\ and\ \bibinfo {author} {\bibfnamefont {M.}~\bibnamefont
  {Lewenstein}},\ }\bibfield  {title} {\bibinfo {title} {Entanglement
  percolation in quantum~networks},\ }\href {https://doi.org/10.1038/nphys549}
  {\bibfield  {journal} {\bibinfo  {journal} {Nature Physics}\ }\textbf
  {\bibinfo {volume} {3}},\ \bibinfo {pages} {256} (\bibinfo {year}
  {2007})}\BibitemShut {NoStop}%
\bibitem [{\citenamefont {Cuquet}\ and\ \citenamefont
  {Calsamiglia}(2009)}]{Cuquet2009}%
  \BibitemOpen
  \bibfield  {author} {\bibinfo {author} {\bibfnamefont {M.}~\bibnamefont
  {Cuquet}}\ and\ \bibinfo {author} {\bibfnamefont {J.}~\bibnamefont
  {Calsamiglia}},\ }\bibfield  {title} {\bibinfo {title} {Entanglement
  percolation in quantum complex networks},\ }\href
  {https://doi.org/10.1103/physrevlett.103.240503} {\bibfield  {journal}
  {\bibinfo  {journal} {Phys. Rev. Lett.}\ }\textbf {\bibinfo {volume} {103}},\
  \bibinfo {pages} {240503} (\bibinfo {year} {2009})}\BibitemShut {NoStop}%
\bibitem [{\citenamefont {Perseguers}\ \emph {et~al.}(2010)\citenamefont
  {Perseguers}, \citenamefont {Lewenstein}, \citenamefont {Ac{\'{\i}}n},\ and\
  \citenamefont {Cirac}}]{Perseguers2010}%
  \BibitemOpen
  \bibfield  {author} {\bibinfo {author} {\bibfnamefont {S.}~\bibnamefont
  {Perseguers}}, \bibinfo {author} {\bibfnamefont {M.}~\bibnamefont
  {Lewenstein}}, \bibinfo {author} {\bibfnamefont {A.}~\bibnamefont
  {Ac{\'{\i}}n}},\ and\ \bibinfo {author} {\bibfnamefont {J.~I.}\ \bibnamefont
  {Cirac}},\ }\bibfield  {title} {\bibinfo {title} {Quantum random networks},\
  }\href {https://doi.org/10.1038/nphys1665} {\bibfield  {journal} {\bibinfo
  {journal} {Nature Physics}\ }\textbf {\bibinfo {volume} {6}},\ \bibinfo
  {pages} {539} (\bibinfo {year} {2010})}\BibitemShut {NoStop}%
\bibitem [{\citenamefont {Faccin}\ \emph {et~al.}(2013)\citenamefont {Faccin},
  \citenamefont {Johnson}, \citenamefont {Biamonte}, \citenamefont {Kais},\
  and\ \citenamefont {Migda{\l}}}]{Faccin2013}%
  \BibitemOpen
  \bibfield  {author} {\bibinfo {author} {\bibfnamefont {M.}~\bibnamefont
  {Faccin}}, \bibinfo {author} {\bibfnamefont {T.}~\bibnamefont {Johnson}},
  \bibinfo {author} {\bibfnamefont {J.}~\bibnamefont {Biamonte}}, \bibinfo
  {author} {\bibfnamefont {S.}~\bibnamefont {Kais}},\ and\ \bibinfo {author}
  {\bibfnamefont {P.}~\bibnamefont {Migda{\l}}},\ }\bibfield  {title} {\bibinfo
  {title} {{Degree Distribution in Quantum Walks on Complex Networks}},\ }\href
  {https://doi.org/10.1103/PhysRevX.3.041007} {\bibfield  {journal} {\bibinfo
  {journal} {Phys. Rev. X}\ }\textbf {\bibinfo {volume} {3}},\ \bibinfo {pages}
  {041007} (\bibinfo {year} {2013})}\BibitemShut {NoStop}%
\bibitem [{\citenamefont {Paparo}\ and\ \citenamefont
  {Martin-Delgado}(2012)}]{Paparo2012}%
  \BibitemOpen
  \bibfield  {author} {\bibinfo {author} {\bibfnamefont {G.~D.}\ \bibnamefont
  {Paparo}}\ and\ \bibinfo {author} {\bibfnamefont {M.~A.}\ \bibnamefont
  {Martin-Delgado}},\ }\bibfield  {title} {\bibinfo {title} {Google in a
  quantum network},\ }\href {https://doi.org/10.1038/srep00444} {\bibfield
  {journal} {\bibinfo  {journal} {Sci. Rep.}\ }\textbf {\bibinfo {volume}
  {2}},\ \bibinfo {pages} {444} (\bibinfo {year} {2012})}\BibitemShut {NoStop}%
\bibitem [{\citenamefont {S{\'{a}}nchez-Burillo}\ \emph
  {et~al.}(2012)\citenamefont {S{\'{a}}nchez-Burillo}, \citenamefont {Duch},
  \citenamefont {G{\'{o}}mez-Garde{\~{n}}es},\ and\ \citenamefont
  {Zueco}}]{SanchezBurillo2012}%
  \BibitemOpen
  \bibfield  {author} {\bibinfo {author} {\bibfnamefont {E.}~\bibnamefont
  {S{\'{a}}nchez-Burillo}}, \bibinfo {author} {\bibfnamefont {J.}~\bibnamefont
  {Duch}}, \bibinfo {author} {\bibfnamefont {J.}~\bibnamefont
  {G{\'{o}}mez-Garde{\~{n}}es}},\ and\ \bibinfo {author} {\bibfnamefont
  {D.}~\bibnamefont {Zueco}},\ }\bibfield  {title} {\bibinfo {title} {Quantum
  navigation and ranking in complex networks},\ }\href
  {https://doi.org/10.1038/srep00605} {\bibfield  {journal} {\bibinfo
  {journal} {Sci. Rep.}\ }\textbf {\bibinfo {volume} {2}},\ \bibinfo {pages}
  {605} (\bibinfo {year} {2012})}\BibitemShut {NoStop}%
\bibitem [{\citenamefont {Faccin}\ \emph {et~al.}(2014)\citenamefont {Faccin},
  \citenamefont {Migdal}, \citenamefont {Johnson}, \citenamefont {Bergholm},\
  and\ \citenamefont {Biamonte}}]{Faccin2014}%
  \BibitemOpen
  \bibfield  {author} {\bibinfo {author} {\bibfnamefont {M.}~\bibnamefont
  {Faccin}}, \bibinfo {author} {\bibfnamefont {P.}~\bibnamefont {Migdal}},
  \bibinfo {author} {\bibfnamefont {T.~H.}\ \bibnamefont {Johnson}}, \bibinfo
  {author} {\bibfnamefont {V.}~\bibnamefont {Bergholm}},\ and\ \bibinfo
  {author} {\bibfnamefont {J.~D.}\ \bibnamefont {Biamonte}},\ }\bibfield
  {title} {\bibinfo {title} {{Community detection in quantum complex
  networks}},\ }\href {https://doi.org/10.1103/PhysRevX.4.041012} {\bibfield
  {journal} {\bibinfo  {journal} {Phys. Rev. X}\ }\textbf {\bibinfo {volume}
  {4}},\ \bibinfo {pages} {1} (\bibinfo {year} {2014})}\BibitemShut {NoStop}%
\bibitem [{\citenamefont {Chakraborty}\ \emph {et~al.}(2016)\citenamefont
  {Chakraborty}, \citenamefont {Novo}, \citenamefont {Ambainis},\ and\
  \citenamefont {Omar}}]{omar16}%
  \BibitemOpen
  \bibfield  {author} {\bibinfo {author} {\bibfnamefont {S.}~\bibnamefont
  {Chakraborty}}, \bibinfo {author} {\bibfnamefont {L.}~\bibnamefont {Novo}},
  \bibinfo {author} {\bibfnamefont {A.}~\bibnamefont {Ambainis}},\ and\
  \bibinfo {author} {\bibfnamefont {Y.}~\bibnamefont {Omar}},\ }\bibfield
  {title} {\bibinfo {title} {{Spatial Search by Quantum Walk is Optimal for
  Almost all Graphs}},\ }\href {https://doi.org/10.1103/PhysRevLett.116.100501}
  {\bibfield  {journal} {\bibinfo  {journal} {Phys. Rev. Lett.}\ }\textbf
  {\bibinfo {volume} {116}},\ \bibinfo {pages} {100501} (\bibinfo {year}
  {2016})}\BibitemShut {NoStop}%
\bibitem [{\citenamefont {M\"ulken}\ \emph {et~al.}(2016)\citenamefont
  {M\"ulken}, \citenamefont {Dolgushev},\ and\ \citenamefont
  {Galiceanu}}]{Mulken2016}%
  \BibitemOpen
  \bibfield  {author} {\bibinfo {author} {\bibfnamefont {O.}~\bibnamefont
  {M\"ulken}}, \bibinfo {author} {\bibfnamefont {M.}~\bibnamefont
  {Dolgushev}},\ and\ \bibinfo {author} {\bibfnamefont {M.}~\bibnamefont
  {Galiceanu}},\ }\bibfield  {title} {\bibinfo {title} {Complex quantum
  networks: From universal breakdown to optimal transport},\ }\href
  {https://doi.org/10.1103/PhysRevE.93.022304} {\bibfield  {journal} {\bibinfo
  {journal} {Phys. Rev. E}\ }\textbf {\bibinfo {volume} {93}},\ \bibinfo
  {pages} {022304} (\bibinfo {year} {2016})}\BibitemShut {NoStop}%
\bibitem [{\citenamefont {Cabot}\ \emph {et~al.}(2018)\citenamefont {Cabot},
  \citenamefont {Galve}, \citenamefont {Egu{\'{\i}}luz}, \citenamefont {Klemm},
  \citenamefont {Maniscalco},\ and\ \citenamefont {Zambrini}}]{Cabot2018}%
  \BibitemOpen
  \bibfield  {author} {\bibinfo {author} {\bibfnamefont {A.}~\bibnamefont
  {Cabot}}, \bibinfo {author} {\bibfnamefont {F.}~\bibnamefont {Galve}},
  \bibinfo {author} {\bibfnamefont {V.~M.}\ \bibnamefont {Egu{\'{\i}}luz}},
  \bibinfo {author} {\bibfnamefont {K.}~\bibnamefont {Klemm}}, \bibinfo
  {author} {\bibfnamefont {S.}~\bibnamefont {Maniscalco}},\ and\ \bibinfo
  {author} {\bibfnamefont {R.}~\bibnamefont {Zambrini}},\ }\bibfield  {title}
  {\bibinfo {title} {Unveiling noiseless clusters in complex quantum
  networks},\ }\href {https://doi.org/10.1038/s41534-018-0108-9} {\bibfield
  {journal} {\bibinfo  {journal} {npj Quantum Inf.}\ }\textbf {\bibinfo
  {volume} {4}},\ \bibinfo {pages} {57} (\bibinfo {year} {2018})}\BibitemShut
  {NoStop}%
\bibitem [{\citenamefont {Nokkala}\ \emph {et~al.}(2016)\citenamefont
  {Nokkala}, \citenamefont {Galve}, \citenamefont {Zambrini}, \citenamefont
  {Maniscalco},\ and\ \citenamefont {Piilo}}]{Nokkala2016}%
  \BibitemOpen
  \bibfield  {author} {\bibinfo {author} {\bibfnamefont {J.}~\bibnamefont
  {Nokkala}}, \bibinfo {author} {\bibfnamefont {F.}~\bibnamefont {Galve}},
  \bibinfo {author} {\bibfnamefont {R.}~\bibnamefont {Zambrini}}, \bibinfo
  {author} {\bibfnamefont {S.}~\bibnamefont {Maniscalco}},\ and\ \bibinfo
  {author} {\bibfnamefont {J.}~\bibnamefont {Piilo}},\ }\bibfield  {title}
  {\bibinfo {title} {{Complex quantum networks as structured environments:
  engineering and probing}},\ }\href {https://doi.org/10.1038/srep26861}
  {\bibfield  {journal} {\bibinfo  {journal} {Sci. Rep.}\ }\textbf {\bibinfo
  {volume} {6}},\ \bibinfo {pages} {26861} (\bibinfo {year}
  {2016})}\BibitemShut {NoStop}%
\bibitem [{\citenamefont {Nokkala}\ \emph
  {et~al.}(2018{\natexlab{a}})\citenamefont {Nokkala}, \citenamefont {Arzani},
  \citenamefont {Galve}, \citenamefont {Zambrini}, \citenamefont {Maniscalco},
  \citenamefont {Piilo}, \citenamefont {Treps},\ and\ \citenamefont
  {Parigi}}]{Nokkala2018}%
  \BibitemOpen
  \bibfield  {author} {\bibinfo {author} {\bibfnamefont {J.}~\bibnamefont
  {Nokkala}}, \bibinfo {author} {\bibfnamefont {F.}~\bibnamefont {Arzani}},
  \bibinfo {author} {\bibfnamefont {F.}~\bibnamefont {Galve}}, \bibinfo
  {author} {\bibfnamefont {R.}~\bibnamefont {Zambrini}}, \bibinfo {author}
  {\bibfnamefont {S.}~\bibnamefont {Maniscalco}}, \bibinfo {author}
  {\bibfnamefont {J.}~\bibnamefont {Piilo}}, \bibinfo {author} {\bibfnamefont
  {N.}~\bibnamefont {Treps}},\ and\ \bibinfo {author} {\bibfnamefont
  {V.}~\bibnamefont {Parigi}},\ }\bibfield  {title} {\bibinfo {title}
  {{Reconfigurable optical implementation of quantum complex networks}},\
  }\href {https://doi.org/10.1088/1367-2630/aabc77} {\bibfield  {journal}
  {\bibinfo  {journal} {New J. Phys.}\ }\textbf {\bibinfo {volume} {20}},\
  \bibinfo {pages} {053024} (\bibinfo {year} {2018}{\natexlab{a}})}\BibitemShut
  {NoStop}%
\bibitem [{\citenamefont {Nokkala}\ \emph
  {et~al.}(2018{\natexlab{b}})\citenamefont {Nokkala}, \citenamefont
  {Maniscalco},\ and\ \citenamefont {Piilo}}]{Nokkala2018a}%
  \BibitemOpen
  \bibfield  {author} {\bibinfo {author} {\bibfnamefont {J.}~\bibnamefont
  {Nokkala}}, \bibinfo {author} {\bibfnamefont {S.}~\bibnamefont
  {Maniscalco}},\ and\ \bibinfo {author} {\bibfnamefont {J.}~\bibnamefont
  {Piilo}},\ }\bibfield  {title} {\bibinfo {title} {{Local probe for
  connectivity and coupling strength in quantum complex networks}},\ }\href
  {https://doi.org/10.1038/s41598-018-30863-2} {\bibfield  {journal} {\bibinfo
  {journal} {Sci. Rep.}\ }\textbf {\bibinfo {volume} {8}},\ \bibinfo {pages}
  {13010} (\bibinfo {year} {2018}{\natexlab{b}})}\BibitemShut {NoStop}%
\bibitem [{\citenamefont {Valdez}\ \emph {et~al.}(2017)\citenamefont {Valdez},
  \citenamefont {Jaschke}, \citenamefont {Vargas},\ and\ \citenamefont
  {Carr}}]{Valdez2017}%
  \BibitemOpen
  \bibfield  {author} {\bibinfo {author} {\bibfnamefont {M.~A.}\ \bibnamefont
  {Valdez}}, \bibinfo {author} {\bibfnamefont {D.}~\bibnamefont {Jaschke}},
  \bibinfo {author} {\bibfnamefont {D.~L.}\ \bibnamefont {Vargas}},\ and\
  \bibinfo {author} {\bibfnamefont {L.~D.}\ \bibnamefont {Carr}},\ }\bibfield
  {title} {\bibinfo {title} {Quantifying complexity in quantum phase
  transitions via mutual information complex networks},\ }\href
  {https://doi.org/10.1103/PhysRevLett.119.225301} {\bibfield  {journal}
  {\bibinfo  {journal} {Phys. Rev. Lett.}\ }\textbf {\bibinfo {volume} {119}},\
  \bibinfo {pages} {225301} (\bibinfo {year} {2017})}\BibitemShut {NoStop}%
\bibitem [{\citenamefont {Sundar}\ \emph {et~al.}(2018)\citenamefont {Sundar},
  \citenamefont {Valdez}, \citenamefont {Carr},\ and\ \citenamefont
  {Hazzard}}]{Bhuvanesh2018}%
  \BibitemOpen
  \bibfield  {author} {\bibinfo {author} {\bibfnamefont {B.}~\bibnamefont
  {Sundar}}, \bibinfo {author} {\bibfnamefont {M.~A.}\ \bibnamefont {Valdez}},
  \bibinfo {author} {\bibfnamefont {L.~D.}\ \bibnamefont {Carr}},\ and\
  \bibinfo {author} {\bibfnamefont {K.~R.~A.}\ \bibnamefont {Hazzard}},\
  }\bibfield  {title} {\bibinfo {title} {Complex-network description of thermal
  quantum states in the ising spin chain},\ }\href
  {https://doi.org/10.1103/PhysRevA.97.052320} {\bibfield  {journal} {\bibinfo
  {journal} {Phys. Rev. A}\ }\textbf {\bibinfo {volume} {97}},\ \bibinfo
  {pages} {052320} (\bibinfo {year} {2018})}\BibitemShut {NoStop}%
\bibitem [{\citenamefont {Nielsen}(2006)}]{Nielsen2006}%
  \BibitemOpen
  \bibfield  {author} {\bibinfo {author} {\bibfnamefont {M.~A.}\ \bibnamefont
  {Nielsen}},\ }\bibfield  {title} {\bibinfo {title} {{Cluster-state quantum
  computation}},\ }\href {https://doi.org/10.1016/S0034-4877(06)80014-5}
  {\bibfield  {journal} {\bibinfo  {journal} {Rep. Math. Phys.}\ }\textbf
  {\bibinfo {volume} {57}},\ \bibinfo {pages} {147} (\bibinfo {year}
  {2006})}\BibitemShut {NoStop}%
\bibitem [{\citenamefont {{De Chiara}}\ and\ \citenamefont
  {Sanpera}(2018)}]{DeChiara2018}%
  \BibitemOpen
  \bibfield  {author} {\bibinfo {author} {\bibfnamefont {G.}~\bibnamefont {{De
  Chiara}}}\ and\ \bibinfo {author} {\bibfnamefont {A.}~\bibnamefont
  {Sanpera}},\ }\bibfield  {title} {\bibinfo {title} {{Genuine quantum
  correlations in quantum many-body systems: a review of recent progress}},\
  }\href {https://doi.org/10.1088/1361-6633/aabf61} {\bibfield  {journal}
  {\bibinfo  {journal} {Rep. Prog. Phys.}\ }\textbf {\bibinfo {volume} {81}},\
  \bibinfo {pages} {074002} (\bibinfo {year} {2018})}\BibitemShut {NoStop}%
\bibitem [{\citenamefont {Amico}\ \emph {et~al.}(2008)\citenamefont {Amico},
  \citenamefont {Fazio}, \citenamefont {Osterloh},\ and\ \citenamefont
  {Vedral}}]{Amico2008}%
  \BibitemOpen
  \bibfield  {author} {\bibinfo {author} {\bibfnamefont {L.}~\bibnamefont
  {Amico}}, \bibinfo {author} {\bibfnamefont {R.}~\bibnamefont {Fazio}},
  \bibinfo {author} {\bibfnamefont {A.}~\bibnamefont {Osterloh}},\ and\
  \bibinfo {author} {\bibfnamefont {V.}~\bibnamefont {Vedral}},\ }\bibfield
  {title} {\bibinfo {title} {{Entanglement in many-body systems}},\ }\href
  {https://doi.org/10.1103/RevModPhys.80.517} {\bibfield  {journal} {\bibinfo
  {journal} {Rev. Mod. Phys.}\ }\textbf {\bibinfo {volume} {80}},\ \bibinfo
  {pages} {517} (\bibinfo {year} {2008})}\BibitemShut {NoStop}%
\bibitem [{\citenamefont {Cotler}\ and\ \citenamefont
  {Wilczek}(2020)}]{cotler2019quantum}%
  \BibitemOpen
  \bibfield  {author} {\bibinfo {author} {\bibfnamefont {J.}~\bibnamefont
  {Cotler}}\ and\ \bibinfo {author} {\bibfnamefont {F.}~\bibnamefont
  {Wilczek}},\ }\bibfield  {title} {\bibinfo {title} {Quantum overlapping
  tomography},\ }\href {https://doi.org/10.1103/PhysRevLett.124.100401}
  {\bibfield  {journal} {\bibinfo  {journal} {Phys. Rev. Lett.}\ }\textbf
  {\bibinfo {volume} {124}},\ \bibinfo {pages} {100401} (\bibinfo {year}
  {2020})}\BibitemShut {NoStop}%
\bibitem [{\citenamefont {Garc\'{\i}a-P\'erez}\ \emph
  {et~al.}(2020)\citenamefont {Garc\'{\i}a-P\'erez}, \citenamefont {Rossi},
  \citenamefont {Sokolov}, \citenamefont {Borrelli},\ and\ \citenamefont
  {Maniscalco}}]{pairwisetomo}%
  \BibitemOpen
  \bibfield  {author} {\bibinfo {author} {\bibfnamefont {G.}~\bibnamefont
  {Garc\'{\i}a-P\'erez}}, \bibinfo {author} {\bibfnamefont {M.~A.~C.}\
  \bibnamefont {Rossi}}, \bibinfo {author} {\bibfnamefont {B.}~\bibnamefont
  {Sokolov}}, \bibinfo {author} {\bibfnamefont {E.-M.}\ \bibnamefont
  {Borrelli}},\ and\ \bibinfo {author} {\bibfnamefont {S.}~\bibnamefont
  {Maniscalco}},\ }\bibfield  {title} {\bibinfo {title} {Pairwise tomography
  networks for many-body quantum systems},\ }\href
  {https://doi.org/10.1103/PhysRevResearch.2.023393} {\bibfield  {journal}
  {\bibinfo  {journal} {Phys. Rev. Research}\ }\textbf {\bibinfo {volume}
  {2}},\ \bibinfo {pages} {023393} (\bibinfo {year} {2020})}\BibitemShut
  {NoStop}%
\bibitem [{\citenamefont {Bonet-Monroig}\ \emph {et~al.}(2019)\citenamefont
  {Bonet-Monroig}, \citenamefont {Babbush},\ and\ \citenamefont
  {O'Brien}}]{bonetmonroig2019nearly}%
  \BibitemOpen
  \bibfield  {author} {\bibinfo {author} {\bibfnamefont {X.}~\bibnamefont
  {Bonet-Monroig}}, \bibinfo {author} {\bibfnamefont {R.}~\bibnamefont
  {Babbush}},\ and\ \bibinfo {author} {\bibfnamefont {T.~E.}\ \bibnamefont
  {O'Brien}},\ }\href@noop {} {\bibinfo {title} {Nearly optimal measurement
  scheduling for partial tomography of quantum states}} (\bibinfo {year}
  {2019}),\ \Eprint {https://arxiv.org/abs/1908.05628} {arXiv:1908.05628
  [quant-ph]} \BibitemShut {NoStop}%
\bibitem [{\citenamefont {Barrat}\ \emph {et~al.}(2008)\citenamefont {Barrat},
  \citenamefont {Barth{\'{e}}lemy},\ and\ \citenamefont
  {Vespignani}}]{Barrat:2008}%
  \BibitemOpen
  \bibfield  {author} {\bibinfo {author} {\bibfnamefont {A.}~\bibnamefont
  {Barrat}}, \bibinfo {author} {\bibfnamefont {M.}~\bibnamefont
  {Barth{\'{e}}lemy}},\ and\ \bibinfo {author} {\bibfnamefont {A.}~\bibnamefont
  {Vespignani}},\ }\href@noop {} {\emph {\bibinfo {title} {Dynamical Porcesses
  on Complex Networks}}}\ (\bibinfo  {publisher} {Cambridge University Press},\
  \bibinfo {address} {Cambridge},\ \bibinfo {year} {2008})\BibitemShut
  {NoStop}%
\bibitem [{\citenamefont {Dorogovtsev}\ \emph {et~al.}(2008)\citenamefont
  {Dorogovtsev}, \citenamefont {Goltsev},\ and\ \citenamefont
  {Mendes}}]{Dorogovtsev:2008}%
  \BibitemOpen
  \bibfield  {author} {\bibinfo {author} {\bibfnamefont {S.~N.}\ \bibnamefont
  {Dorogovtsev}}, \bibinfo {author} {\bibfnamefont {A.~V.}\ \bibnamefont
  {Goltsev}},\ and\ \bibinfo {author} {\bibfnamefont {J.~F.~F.}\ \bibnamefont
  {Mendes}},\ }\bibfield  {title} {\bibinfo {title} {Critical phenomena in
  complex networks},\ }\href {https://doi.org/10.1103/RevModPhys.80.1275}
  {\bibfield  {journal} {\bibinfo  {journal} {Rev. Mod. Phys.}\ }\textbf
  {\bibinfo {volume} {80}},\ \bibinfo {pages} {1275} (\bibinfo {year}
  {2008})}\BibitemShut {NoStop}%
\bibitem [{\citenamefont {Newman}(2010)}]{Newman:2010}%
  \BibitemOpen
  \bibfield  {author} {\bibinfo {author} {\bibfnamefont {M.}~\bibnamefont
  {Newman}},\ }\href@noop {} {\emph {\bibinfo {title} {Networks: An
  Introduction}}}\ (\bibinfo  {publisher} {Oxford University Press, Inc.},\
  \bibinfo {address} {New York, NY, USA},\ \bibinfo {year} {2010})\BibitemShut
  {NoStop}%
\bibitem [{\citenamefont {Bogu\~{n}\'{a}}\ \emph {et~al.}(2010)\citenamefont
  {Bogu\~{n}\'{a}}, \citenamefont {Papadopoulos},\ and\ \citenamefont
  {Krioukov}}]{Boguna:2010}%
  \BibitemOpen
  \bibfield  {author} {\bibinfo {author} {\bibfnamefont {M.}~\bibnamefont
  {Bogu\~{n}\'{a}}}, \bibinfo {author} {\bibfnamefont {F.}~\bibnamefont
  {Papadopoulos}},\ and\ \bibinfo {author} {\bibfnamefont {D.}~\bibnamefont
  {Krioukov}},\ }\bibfield  {title} {\bibinfo {title} {{Sustaining the Internet
  with Hyperbolic Mapping}},\ }\href {https://doi.org/10.1038/ncomms1063}
  {\bibfield  {journal} {\bibinfo  {journal} {Nature Comms}\ }\textbf {\bibinfo
  {volume} {1}},\ \bibinfo {pages} {62} (\bibinfo {year} {2010})}\BibitemShut
  {NoStop}%
\bibitem [{\citenamefont {Sporns}\ \emph {et~al.}(2005)\citenamefont {Sporns},
  \citenamefont {Tononi},\ and\ \citenamefont {Kötter}}]{Sporns:2005}%
  \BibitemOpen
  \bibfield  {author} {\bibinfo {author} {\bibfnamefont {O.}~\bibnamefont
  {Sporns}}, \bibinfo {author} {\bibfnamefont {G.}~\bibnamefont {Tononi}},\
  and\ \bibinfo {author} {\bibfnamefont {R.}~\bibnamefont {Kötter}},\
  }\bibfield  {title} {\bibinfo {title} {The human connectome: A structural
  description of the human brain},\ }\href
  {https://doi.org/10.1371/journal.pcbi.0010042} {\bibfield  {journal}
  {\bibinfo  {journal} {PLOS Computational Biology}\ }\textbf {\bibinfo
  {volume} {1}} (\bibinfo {year} {2005})}\BibitemShut {NoStop}%
\bibitem [{\citenamefont {Allard}\ and\ \citenamefont
  {Serrano}(2020)}]{Allard:2020}%
  \BibitemOpen
  \bibfield  {author} {\bibinfo {author} {\bibfnamefont {A.}~\bibnamefont
  {Allard}}\ and\ \bibinfo {author} {\bibfnamefont {M.~A.}\ \bibnamefont
  {Serrano}},\ }\bibfield  {title} {\bibinfo {title} {Navigable maps of
  structural brain networks across species},\ }\href
  {https://doi.org/10.1371/journal.pcbi.1007584} {\bibfield  {journal}
  {\bibinfo  {journal} {PLOS Computational Biology}\ }\textbf {\bibinfo
  {volume} {16}},\ \bibinfo {pages} {1} (\bibinfo {year} {2020})}\BibitemShut
  {NoStop}%
\bibitem [{\citenamefont {Kleineberg}\ and\ \citenamefont
  {Bogu\~n\'a}(2014)}]{Kleineberg:2014}%
  \BibitemOpen
  \bibfield  {author} {\bibinfo {author} {\bibfnamefont {K.-K.}\ \bibnamefont
  {Kleineberg}}\ and\ \bibinfo {author} {\bibfnamefont {M.}~\bibnamefont
  {Bogu\~n\'a}},\ }\bibfield  {title} {\bibinfo {title} {Evolution of the
  digital society reveals balance between viral and mass media influence},\
  }\href {https://doi.org/10.1103/PhysRevX.4.031046} {\bibfield  {journal}
  {\bibinfo  {journal} {Phys. Rev. X}\ }\textbf {\bibinfo {volume} {4}},\
  \bibinfo {pages} {031046} (\bibinfo {year} {2014})}\BibitemShut {NoStop}%
\bibitem [{\citenamefont {Garc{\'i}a-P{\'e}rez}\ \emph
  {et~al.}(2018)\citenamefont {Garc{\'i}a-P{\'e}rez}, \citenamefont
  {Bogu{\~{n}}{\'a}},\ and\ \citenamefont {Serrano}}]{Garcia-Perez:2018}%
  \BibitemOpen
  \bibfield  {author} {\bibinfo {author} {\bibfnamefont {G.}~\bibnamefont
  {Garc{\'i}a-P{\'e}rez}}, \bibinfo {author} {\bibfnamefont {M.}~\bibnamefont
  {Bogu{\~{n}}{\'a}}},\ and\ \bibinfo {author} {\bibfnamefont {M.~{\'A}.}\
  \bibnamefont {Serrano}},\ }\bibfield  {title} {\bibinfo {title} {Multiscale
  unfolding of real networks by geometric renormalization},\ }\href
  {https://doi.org/10.1038/s41567-018-0072-5} {\bibfield  {journal} {\bibinfo
  {journal} {Nature Physics}\ }\textbf {\bibinfo {volume} {14}},\ \bibinfo
  {pages} {583} (\bibinfo {year} {2018})}\BibitemShut {NoStop}%
\bibitem [{\citenamefont {Bogu\~n\'a}\ \emph {et~al.}(2003)\citenamefont
  {Bogu\~n\'a}, \citenamefont {Pastor-Satorras},\ and\ \citenamefont
  {Vespignani}}]{Boguna:2003}%
  \BibitemOpen
  \bibfield  {author} {\bibinfo {author} {\bibfnamefont {M.}~\bibnamefont
  {Bogu\~n\'a}}, \bibinfo {author} {\bibfnamefont {R.}~\bibnamefont
  {Pastor-Satorras}},\ and\ \bibinfo {author} {\bibfnamefont {A.}~\bibnamefont
  {Vespignani}},\ }\bibfield  {title} {\bibinfo {title} {Absence of epidemic
  threshold in scale-free networks with degree correlations},\ }\href
  {https://doi.org/10.1103/PhysRevLett.90.028701} {\bibfield  {journal}
  {\bibinfo  {journal} {Phys. Rev. Lett.}\ }\textbf {\bibinfo {volume} {90}},\
  \bibinfo {pages} {028701} (\bibinfo {year} {2003})}\BibitemShut {NoStop}%
\bibitem [{\citenamefont {Serrano}\ \emph {et~al.}(2009)\citenamefont
  {Serrano}, \citenamefont {Bogu{\~n}{\'a}},\ and\ \citenamefont
  {Vespignani}}]{Serrano6483}%
  \BibitemOpen
  \bibfield  {author} {\bibinfo {author} {\bibfnamefont {M.~{\'A}.}\
  \bibnamefont {Serrano}}, \bibinfo {author} {\bibfnamefont {M.}~\bibnamefont
  {Bogu{\~n}{\'a}}},\ and\ \bibinfo {author} {\bibfnamefont {A.}~\bibnamefont
  {Vespignani}},\ }\bibfield  {title} {\bibinfo {title} {Extracting the
  multiscale backbone of complex weighted networks},\ }\href
  {https://doi.org/10.1073/pnas.0808904106} {\bibfield  {journal} {\bibinfo
  {journal} {Proc. Natl. Acad. Sci.}\ }\textbf {\bibinfo {volume} {106}},\
  \bibinfo {pages} {6483} (\bibinfo {year} {2009})}\BibitemShut {NoStop}%
\bibitem [{\citenamefont {Onnela}\ \emph {et~al.}(2005)\citenamefont {Onnela},
  \citenamefont {Saram\"aki}, \citenamefont {Kert\'esz},\ and\ \citenamefont
  {Kaski}}]{Onnela2005}%
  \BibitemOpen
  \bibfield  {author} {\bibinfo {author} {\bibfnamefont {J.-P.}\ \bibnamefont
  {Onnela}}, \bibinfo {author} {\bibfnamefont {J.}~\bibnamefont {Saram\"aki}},
  \bibinfo {author} {\bibfnamefont {J.}~\bibnamefont {Kert\'esz}},\ and\
  \bibinfo {author} {\bibfnamefont {K.}~\bibnamefont {Kaski}},\ }\bibfield
  {title} {\bibinfo {title} {Intensity and coherence of motifs in weighted
  complex networks},\ }\href {https://doi.org/10.1103/PhysRevE.71.065103}
  {\bibfield  {journal} {\bibinfo  {journal} {Phys. Rev. E}\ }\textbf {\bibinfo
  {volume} {71}},\ \bibinfo {pages} {065103} (\bibinfo {year}
  {2005})}\BibitemShut {NoStop}%
\bibitem [{\citenamefont {Osterloh}\ \emph {et~al.}(2002)\citenamefont
  {Osterloh}, \citenamefont {Amico}, \citenamefont {Falci},\ and\ \citenamefont
  {Fazio}}]{Osterloh2002}%
  \BibitemOpen
  \bibfield  {author} {\bibinfo {author} {\bibfnamefont {A.}~\bibnamefont
  {Osterloh}}, \bibinfo {author} {\bibfnamefont {L.}~\bibnamefont {Amico}},
  \bibinfo {author} {\bibfnamefont {G.}~\bibnamefont {Falci}},\ and\ \bibinfo
  {author} {\bibfnamefont {R.}~\bibnamefont {Fazio}},\ }\bibfield  {title}
  {\bibinfo {title} {{Scaling of entanglement close to a quantum phase
  transition}},\ }\href {https://doi.org/10.1038/416608a} {\bibfield  {journal}
  {\bibinfo  {journal} {Nature}\ }\textbf {\bibinfo {volume} {416}},\ \bibinfo
  {pages} {608} (\bibinfo {year} {2002})}\BibitemShut {NoStop}%
\bibitem [{\citenamefont {Osborne}\ and\ \citenamefont
  {Nielsen}(2002)}]{Osborne2002}%
  \BibitemOpen
  \bibfield  {author} {\bibinfo {author} {\bibfnamefont {T.~J.}\ \bibnamefont
  {Osborne}}\ and\ \bibinfo {author} {\bibfnamefont {M.~A.}\ \bibnamefont
  {Nielsen}},\ }\bibfield  {title} {\bibinfo {title} {{Entanglement in a simple
  quantum phase transition}},\ }\href
  {https://doi.org/10.1103/PhysRevA.66.032110} {\bibfield  {journal} {\bibinfo
  {journal} {Phys. Rev. A}\ }\textbf {\bibinfo {volume} {66}},\ \bibinfo
  {pages} {032110} (\bibinfo {year} {2002})}\BibitemShut {NoStop}%
\bibitem [{\citenamefont {Lanyon}\ \emph {et~al.}(2017)\citenamefont {Lanyon},
  \citenamefont {Maier}, \citenamefont {Holz{\"{a}}pfel}, \citenamefont
  {Baumgratz}, \citenamefont {Hempel}, \citenamefont {Jurcevic}, \citenamefont
  {Dhand}, \citenamefont {Buyskikh}, \citenamefont {Daley}, \citenamefont
  {Cramer}, \citenamefont {Plenio}, \citenamefont {Blatt},\ and\ \citenamefont
  {Roos}}]{Lanyon2017}%
  \BibitemOpen
  \bibfield  {author} {\bibinfo {author} {\bibfnamefont {B.~P.}\ \bibnamefont
  {Lanyon}}, \bibinfo {author} {\bibfnamefont {C.}~\bibnamefont {Maier}},
  \bibinfo {author} {\bibfnamefont {M.}~\bibnamefont {Holz{\"{a}}pfel}},
  \bibinfo {author} {\bibfnamefont {T.}~\bibnamefont {Baumgratz}}, \bibinfo
  {author} {\bibfnamefont {C.}~\bibnamefont {Hempel}}, \bibinfo {author}
  {\bibfnamefont {P.}~\bibnamefont {Jurcevic}}, \bibinfo {author}
  {\bibfnamefont {I.}~\bibnamefont {Dhand}}, \bibinfo {author} {\bibfnamefont
  {A.~S.}\ \bibnamefont {Buyskikh}}, \bibinfo {author} {\bibfnamefont {A.~J.}\
  \bibnamefont {Daley}}, \bibinfo {author} {\bibfnamefont {M.}~\bibnamefont
  {Cramer}}, \bibinfo {author} {\bibfnamefont {M.~B.}\ \bibnamefont {Plenio}},
  \bibinfo {author} {\bibfnamefont {R.}~\bibnamefont {Blatt}},\ and\ \bibinfo
  {author} {\bibfnamefont {C.~F.}\ \bibnamefont {Roos}},\ }\bibfield  {title}
  {\bibinfo {title} {{Efficient tomography of a quantum many-body system}},\
  }\href {https://doi.org/10.1038/nphys4244} {\bibfield  {journal} {\bibinfo
  {journal} {Nat. Phys.}\ }\textbf {\bibinfo {volume} {13}},\ \bibinfo {pages}
  {1158} (\bibinfo {year} {2017})}\BibitemShut {NoStop}%
\bibitem [{\citenamefont {Gross}\ and\ \citenamefont
  {Bloch}(2017)}]{Gross2017}%
  \BibitemOpen
  \bibfield  {author} {\bibinfo {author} {\bibfnamefont {C.}~\bibnamefont
  {Gross}}\ and\ \bibinfo {author} {\bibfnamefont {I.}~\bibnamefont {Bloch}},\
  }\bibfield  {title} {\bibinfo {title} {{Quantum simulations with ultracold
  atoms in optical lattices}},\ }\href
  {https://doi.org/10.1126/science.aal3837} {\bibfield  {journal} {\bibinfo
  {journal} {Science}\ }\textbf {\bibinfo {volume} {357}},\ \bibinfo {pages}
  {995} (\bibinfo {year} {2017})}\BibitemShut {NoStop}%
\bibitem [{\citenamefont {Son}\ \emph {et~al.}(2009)\citenamefont {Son},
  \citenamefont {Amico}, \citenamefont {Plastina},\ and\ \citenamefont
  {Vedral}}]{Son2009}%
  \BibitemOpen
  \bibfield  {author} {\bibinfo {author} {\bibfnamefont {W.}~\bibnamefont
  {Son}}, \bibinfo {author} {\bibfnamefont {L.}~\bibnamefont {Amico}}, \bibinfo
  {author} {\bibfnamefont {F.}~\bibnamefont {Plastina}},\ and\ \bibinfo
  {author} {\bibfnamefont {V.}~\bibnamefont {Vedral}},\ }\bibfield  {title}
  {\bibinfo {title} {{Quantum instability and edge entanglement in the
  quasi-long-range order}},\ }\href
  {https://doi.org/10.1103/PhysRevA.79.022302} {\bibfield  {journal} {\bibinfo
  {journal} {Phys. Rev. A}\ }\textbf {\bibinfo {volume} {79}},\ \bibinfo
  {pages} {1} (\bibinfo {year} {2009})}\BibitemShut {NoStop}%
\bibitem [{\citenamefont {Lieb}\ \emph {et~al.}(1961)\citenamefont {Lieb},
  \citenamefont {Schultz},\ and\ \citenamefont {Mattis}}]{Lieb1961}%
  \BibitemOpen
  \bibfield  {author} {\bibinfo {author} {\bibfnamefont {E.}~\bibnamefont
  {Lieb}}, \bibinfo {author} {\bibfnamefont {T.}~\bibnamefont {Schultz}},\ and\
  \bibinfo {author} {\bibfnamefont {D.}~\bibnamefont {Mattis}},\ }\bibfield
  {title} {\bibinfo {title} {{Two soluble models of an antiferromagnetic
  chain}},\ }\href {https://doi.org/10.1016/0003-4916(61)90115-4} {\bibfield
  {journal} {\bibinfo  {journal} {Ann. Phys.}\ }\textbf {\bibinfo {volume}
  {16}},\ \bibinfo {pages} {407} (\bibinfo {year} {1961})}\BibitemShut
  {NoStop}%
\bibitem [{\citenamefont {Fortunato}\ and\ \citenamefont
  {Hric}(2016)}]{Fortunato:2016}%
  \BibitemOpen
  \bibfield  {author} {\bibinfo {author} {\bibfnamefont {S.}~\bibnamefont
  {Fortunato}}\ and\ \bibinfo {author} {\bibfnamefont {D.}~\bibnamefont
  {Hric}},\ }\bibfield  {title} {\bibinfo {title} {Community detection in
  networks: A user guide},\ }\href
  {https://doi.org/https://doi.org/10.1016/j.physrep.2016.09.002} {\bibfield
  {journal} {\bibinfo  {journal} {Physics Reports}\ }\textbf {\bibinfo {volume}
  {659}},\ \bibinfo {pages} {1 } (\bibinfo {year} {2016})},\ \bibinfo {note}
  {community detection in networks: A user guide}\BibitemShut {NoStop}%
\bibitem [{\citenamefont {Raghavan}\ \emph {et~al.}(2007)\citenamefont
  {Raghavan}, \citenamefont {Albert},\ and\ \citenamefont
  {Kumara}}]{raghavan2007}%
  \BibitemOpen
  \bibfield  {author} {\bibinfo {author} {\bibfnamefont {U.~N.}\ \bibnamefont
  {Raghavan}}, \bibinfo {author} {\bibfnamefont {R.}~\bibnamefont {Albert}},\
  and\ \bibinfo {author} {\bibfnamefont {S.}~\bibnamefont {Kumara}},\
  }\bibfield  {title} {\bibinfo {title} {Near linear time algorithm to detect
  community structures in large-scale networks},\ }\href
  {https://doi.org/10.1103/PhysRevE.76.036106} {\bibfield  {journal} {\bibinfo
  {journal} {Phys. Rev. E}\ }\textbf {\bibinfo {volume} {76}},\ \bibinfo
  {pages} {036106} (\bibinfo {year} {2007})}\BibitemShut {NoStop}%
\bibitem [{\citenamefont {Fradkin}\ and\ \citenamefont
  {Susskind}(1978)}]{susskind1978}%
  \BibitemOpen
  \bibfield  {author} {\bibinfo {author} {\bibfnamefont {E.}~\bibnamefont
  {Fradkin}}\ and\ \bibinfo {author} {\bibfnamefont {L.}~\bibnamefont
  {Susskind}},\ }\bibfield  {title} {\bibinfo {title} {Order and disorder in
  gauge systems and magnets},\ }\href
  {https://doi.org/10.1103/PhysRevD.17.2637} {\bibfield  {journal} {\bibinfo
  {journal} {Phys. Rev. D}\ }\textbf {\bibinfo {volume} {17}},\ \bibinfo
  {pages} {2637} (\bibinfo {year} {1978})}\BibitemShut {NoStop}%
\bibitem [{\citenamefont {Cordasco}\ and\ \citenamefont
  {Gargano}(2010)}]{cordasco2010}%
  \BibitemOpen
  \bibfield  {author} {\bibinfo {author} {\bibfnamefont {G.}~\bibnamefont
  {Cordasco}}\ and\ \bibinfo {author} {\bibfnamefont {L.}~\bibnamefont
  {Gargano}},\ }\bibfield  {title} {\bibinfo {title} {Community detection via
  semi-synchronous label propagation algorithms},\ }in\ \href@noop {} {\emph
  {\bibinfo {booktitle} {2010 IEEE International Workshop on: Business
  Applications of Social Network Analysis (BASNA)}}}\ (\bibinfo {organization}
  {IEEE},\ \bibinfo {year} {2010})\ pp.\ \bibinfo {pages} {1--8}\BibitemShut
  {NoStop}%
\bibitem [{\citenamefont {Hagberg}\ \emph {et~al.}(2008)\citenamefont
  {Hagberg}, \citenamefont {Schult},\ and\ \citenamefont {Swart}}]{networkx}%
  \BibitemOpen
  \bibfield  {author} {\bibinfo {author} {\bibfnamefont {A.~A.}\ \bibnamefont
  {Hagberg}}, \bibinfo {author} {\bibfnamefont {D.~A.}\ \bibnamefont
  {Schult}},\ and\ \bibinfo {author} {\bibfnamefont {P.~J.}\ \bibnamefont
  {Swart}},\ }\bibfield  {title} {\bibinfo {title} {Exploring network
  structure, dynamics, and function using networkx},\ }in\ \href@noop {} {\emph
  {\bibinfo {booktitle} {Proceedings of the 7th Python in Science
  Conference}}},\ \bibinfo {editor} {edited by\ \bibinfo {editor}
  {\bibfnamefont {G.}~\bibnamefont {Varoquaux}}, \bibinfo {editor}
  {\bibfnamefont {T.}~\bibnamefont {Vaught}},\ and\ \bibinfo {editor}
  {\bibfnamefont {J.}~\bibnamefont {Millman}}}\ (\bibinfo {address} {Pasadena,
  CA USA},\ \bibinfo {year} {2008})\ pp.\ \bibinfo {pages} {11 --
  15}\BibitemShut {NoStop}%
\end{thebibliography}%

\end{document}